\documentclass[acmsmall,screen,nonacm]{acmart}

\usepackage{amsmath}

\usepackage{amssymb}
\usepackage{graphicx}
\usepackage{tikz}
\usepackage{array}
\usepackage{tabularx}
\usepackage{booktabs}
\usepackage{multirow}
\usepackage{makecell}
\usepackage{colortbl}
\usepackage{float}
\usepackage{placeins}
\usepackage{algorithm}
\usepackage{algpseudocode}
\usetikzlibrary{arrows.meta,positioning}
\usepackage{xspace}
\usepackage[normalem]{ulem}
\definecolor{table_shallow_blue}{RGB}{229,238,251}
\usepackage{enumitem}
\newcommand{\method}[1]{\textsc{#1}}

\newcommand{\modify}[1]{{#1}}

\newcommand{\findingbox}[1]{%
  \par\addvspace{6pt}\noindent
  {\setlength{\fboxsep}{6pt}%
    \colorbox{gray!15}{%
      \parbox{\dimexpr\linewidth-2\fboxsep\relax}{\emph{#1}}%
    }%
  }%
  \par\addvspace{6pt}%
}

\newcommand{\spine}{\method{Spine}\xspace}

\newcommand{\spinetree}{SpineTree}
\newcommand{\spinecodex}{\method{SpineCodex}}
\newcommand{\spinepi}{\method{SpinePi}}
\newcommand{\basecodex}{\method{Codex}}
\newcommand{\basepi}{\method{Pi}}
\newcommand{\agentdiet}{\method{AgentDiet}}
\newcommand{\agentfold}{\method{AgentFold}}

\newcommand{\attncompress}{\method{AttnCompress}}
\newcommand{\swepruner}{\method{SWE-Pruner}}
\newcommand{\autocompact}{\method{AutoCompact}}

\newcommand{\contextfolding}{\method{Context Folding}}
\newcommand{\recursivesubagent}{\method{Recursive Subagent}}

\newcommand{\rah}{\method{Recursive Agent Harnesses}}

\newcommand{\codex}{Codex\xspace}
\newcommand{\piagent}{Pi\xspace}
\newcommand{\openai}{OpenAI}

\newcommand{\nvidia}{NVIDIA}

\newcommand{\git}{Git}
\newcommand{\vllm}{vLLM}

\newcommand{\tinkerapi}{Tinker API}
\newcommand{\aOneHundred}{A100}
\newcommand{\hOneHundred}{H100}

\newcommand{\corvus}{\method{CORVUS}}
\newcommand{\swemem}{\method{SWE-MeM}}
\newcommand{\ledger}{\method{Ledger}}
\newcommand{\hiagent}{\method{HiAgent}}
\newcommand{\foldgrpo}{\method{FoldGRPO}}
\newcommand{\thread}{\method{THREAD}}

\newcommand{\agentprune}{\method{AgentPrune}}
\newcommand{\kvcomm}{\method{KVCOMM}}

\newcommand{\claudecode}{Claude Code}

\newcommand{\nushell}{Nushell}

\newcommand{\pipelinedata}{PipelineData}
\newcommand{\ffmpeg}{FFmpeg}
\newcommand{\sqlite}{SQLite}
\newcommand{\postgresql}{PostgreSQL}
\newcommand{\zig}{Zig}
\newcommand{\libswscale}{libswscale}

\newcommand{\swemilestone}{SWE-Milestone}
\newcommand{\programbench}{ProgramBench}
\newcommand{\frontierswe}{FrontierSWE}

\newcommand{\gptfivesixsol}{GPT-5.6 Sol}
\newcommand{\grokfoursix}{Grok 4.6}
\newcommand{\geminithreeeightflash}{Gemini 3.8 Flash}
\newcommand{\qwenthreeeighttwentysevenb}{Qwen 3.8 27B}

\newcommand{\workunit}{work scope}
\newcommand{\workunits}{work scopes}

\newcommand{\parsestack}{ParseStack}
\newcommand{\sjBranchSymbol}{Branch}
\newcommand{\sjReducedNodeSymbol}{TreeNode}
\newcommand{\sjSpawnNodeSymbol}{SpawnNode}

\newcommand{\sjOpenName}{Open}
\newcommand{\sjCloseName}{Close}
\newcommand{\sjNextName}{Next}
\newcommand{\sjSpawnName}{Spawn}
\newcommand{\openop}{\texttt{\sjOpenName}}
\newcommand{\closeop}{\texttt{\sjCloseName}}
\newcommand{\nextop}{\texttt{\sjNextName}}
\newcommand{\spawnop}{\texttt{\sjSpawnName}}

\begin{document}

\title[Spine: Compiling Subtask Trees into Structured Context]{Spine: Compiling Subtask Trees into Structured Context for Long-Horizon Coding Agents}

\author{Jiahong Xiang}
\affiliation{%
  \institution{Southern University of Science and Technology}
  \city{Shenzhen}
  \country{China}}
\email{xiangjh2022@mail.sustech.edu.cn}

\author{Kunqiu Chen}
\affiliation{%
  \institution{Southern University of Science and Technology}
  \city{Shenzhen}
  \country{China}}
\email{11911626@mail.sustech.edu.cn}

\author{Lingming Zhang}
\affiliation{%
  \institution{University of Illinois at Urbana-Champaign}
  \city{Urbana}
  \state{IL}
  \country{USA}}
\email{lingming@illinois.edu}

\author{Yuqun Zhang}
\affiliation{%
  \institution{Southern University of Science and Technology}
  \city{Shenzhen}
  \country{China}}
\email{zhangyq@sustech.edu.cn}

\begin{abstract}
Long-horizon coding tasks contain nested subtasks. Typically, agents record interactions sequentially into a growing history as model context, causing severe context inflation and performance degradation.
One group of existing context-management approaches compresses or trims the history without explicit subtask boundaries, such that context reduction can discard information needed by unfinished subtasks, leading to repeated exploration and disrupted task progress. On the other hand, recursive subagent delegation isolates completed subtasks through child agents via assigning every subtask boundary to delegation, which conflates subtask granularity with executing agent choice, adding execution and coordination overhead. This raises a question: how can model context align with subtask nesting and completion without incurring the high overhead of additional agents? To answer this question, we introduce \spine{}, an interface and runtime that align model context with subtask nesting and completion in a just-in-time manner. Specifically, an agent declares explicit subtask boundaries through \workunits{}, allowing the same agent to execute nested subtasks inline without multi-agent overhead, while reserving delegation for independent parallel tasks. The runtime organizes execution in \spinetree{}, a persistent tree that serves as the structural spine recording subtask relations, local interactions, completion states, and returned results. Before each model call, the runtime compiles the tree into model context \emph{just in time}, retaining unfinished subtasks' interactions and replacing completed subtasks' details with their submitted results, while preserving prefix-cache reuse.

We integrate \spine{} into \codex{}, \openai{}'s popular coding-agent harness, to build \spinecodex{}. We evaluate it on three widely-adopted long-horizon benchmarks: \swemilestone{} for repository evolution, \programbench{} for whole-program reconstruction, and \frontierswe{} for open-ended engineering. With \gptfivesixsol{}, \spinecodex{} improves \swemilestone{} Score by 11.8\% while reducing total cost by 32.4\%, compared with native \codex{}. It also achieves a comparable Score to \recursivesubagent{} at 73.7\% lower total cost. With the same model, it improves the average \programbench{} test-pass rate by 17.3\% compared with native \codex{} and improves Mean@5, i.e., the mean score over five trials per task, on eight of the ten evaluated \frontierswe{} tasks. To evaluate generalizability, we integrate \spine{} into \piagent{}, a popular open-source coding-agent harness with an extensible architecture, through a reusable software package. With \geminithreeeightflash{} and the open-weight \qwenthreeeighttwentysevenb{}, this integration improves \swemilestone{} Score by 9.5--17.5\% and the average \programbench{} test-pass rate by 4.3--8.7\%, compared with native \piagent{}.

\end{abstract}


\maketitle

\section{Introduction}
\label{sec:introduction}

Coding agents are increasingly used for complex, long-horizon software-engineering tasks~\cite{anthropicRakuten,codex2026agenticwork}. For example, in May 2026, roughly one in four sampled active individual users of \codex{}, \openai{}'s coding agent, had submitted a request estimated to require at least eight hours of human work~\cite{codex2026agenticwork}. These tasks contain nested subtasks, i.e., subtasks themselves containing further subtasks. When a subtask completes, the agent uses its result to resume its parent task, i.e., the task that directly contains it. To complete such tasks, a coding agent iteratively reasons, invokes tools, and uses the resulting observations to guide subsequent interactions~\cite{yao2023react}. The agent harness, i.e., the software that runs tools and prepares model inputs, encapsulates these interactions into subsequent model calls as a growing interaction history~\cite{wang2025openhands}, which forms part of the model context as the input information supplied to each model call. However, simply recording interactions in turn as history is impractical, \modify{since it could easily inflate the context and reduce agent effectiveness, making it essential to select which interactions to encapsulate. Intuitively, for completed subtasks, only their returned results are likely needed by their parent task rather than detailed original interactions, which are in fact essential for proceeding with unfinished subtasks~\cite{xiao2026agentdiet}.} %

Existing approaches manage context by compressing or trimming the interaction history, or by assigning subtasks to child agents. History compression and trimming reduce the information encapsulated into later calls. For instance, \agentdiet{} removes redundant or expired content~\cite{xiao2026agentdiet}, while \agentfold{} lets the agent summarize selected parts of the history at different granular levels~\cite{ye2025agentfold}. \modify{\contextfolding{} uses a plan--branch--return workflow trained with \foldgrpo{}, in which the returned result of a completed branch replaces its intermediate interactions}~\cite{sun2025contextfolding}. However, reducing interaction history does not by itself maintain subtask relations, completion status, or where unfinished work should resume. When this information must be recovered from reduced context, the model may lose track of progress, repeat exploration~\cite{deng2026evoclaw}, or declare completion while work remains unfinished~\cite{young2025harnesses}. These failures add execution cost and disrupt continued progress. %

 Recursive subagent delegation instead assigns subtasks to child agents with separate contexts, which can in turn delegate further subtasks and return their results. For example, \rah{} gives each child a complete tool-using harness with its own workspace and context~\cite{rah2026}. \modify{These calls represent subtask nesting and isolate completed work through returned results, but using them to establish every subtask boundary forces delegation and adds execution and coordination overhead, even when the current agent could perform the work itself. Studies of multi-agent systems show that communication inflation and redundant messages can directly degrade task performance~\cite{zhang2025agentprune,zhu2025multiagentbench}, while divergent context prefixes force shared information to be repeatedly reprocessed without KV-cache reuse~\cite{ye2025kvcomm}. Crucially, whether a subtask is delegated to another agent or executed locally by the current agent, the agent still makes decisions about when the subtask starts, completes, and returns. However, existing interaction histories lack an explicit representation of these boundaries and completion states.} %

\modify{To summarize, these observations lead to three issues in context management:
(1)~the lack of an explicit representation of subtask boundaries and completion status in the interaction history;
(2)~the lack of separation between subtask granularity and executing agent choice; %
and (3)~the lack of a history representation distinguishing between completed and unfinished subtasks for parent tasks.}

\modify{This raises a question: how can model context align with subtask nesting and completion without incurring the high overhead of additional agents?}  Ideally, the same agent should be able to complete a subtask and resume its parent task with the result, while retaining information needed by other unfinished subtasks with explicit subtask boundaries. %

To answer this question, we introduce \spine{}, an interface and runtime that align model context with subtask nesting and completion while supporting both local execution and delegation in a just-in-time manner. \modify{Specifically, \spine{} addresses the aforementioned issues through three key designs.
(1)~To provide an explicit representation of subtask boundaries and completion status, the agent declares a \workunit{}, i.e., a subtask with an explicit entry, a parent task, and a result submitted on completion.
(2)~To separate subtask granularity from executing agent choice, a work scope can span multiple model calls and tool interactions, where the current agent can open nested scopes within it, complete each scope by submitting a result, and resume the parent task locally, while independent subtasks can be delegated to other agents for parallel execution so that subtask boundaries do not determine which agent executes the work.
(3)~To distinguish completed subtasks and unfinished subtasks for parent tasks, the runtime maintains a persistent execution tree, namely \spinetree{}, and compiles it into model context \emph{just in time} before each model call, retaining unfinished work scopes' interactions and replacing completed scopes' details with their submitted results.} \modify{More specifically, \spinetree{} serves as the structural spine recording the current agent's scopes, their nesting, interactions, and completion status, together with submitted results and results returned by delegated agents. Completing a work scope therefore both resumes its parent with the submitted result and changes the context supplied to the next call. Notably, the detailed interactions remain preserved in \spinetree{}, while the unchanged context preceding the scope remains eligible for prefix-cache reuse.}

We integrate \spine{} into the agent harness of \codex{}, \openai{}'s widely used coding product~\cite{openai2026codexrepo,openai2026codexknowledgework}, to build \spinecodex{}. We evaluate effectiveness and cost using \gptfivesixsol{}~\cite{openai2026gpt56sol} and \grokfoursix{}~\cite{xai2026grok46} on three widely-adopted long-horizon coding benchmarks: \swemilestone{}~\cite{deng2026evoclaw}, \programbench{}~\cite{yang2026programbench}, and \frontierswe{}~\cite{proximal2026frontierswe}. On \swemilestone{} with \gptfivesixsol{}, \spinecodex{} improves Score by 11.8\%, 16.7\%, and 14.8\% while reducing total cost by 32.4\%, 58.1\%, and 22.7\%, compared with native \codex{}, \agentdiet{}, and \contextfolding{}, respectively. It also achieves a comparable Score to \recursivesubagent{} at 73.7\% lower total cost. With the same model, it improves the average \programbench{} test-pass rate by 17.3\% compared with native \codex{}, and \frontierswe{}'s Mean@5, i.e., the mean score over five trials per task, on eight of ten evaluated tasks. To evaluate \modify{generalizability}, we integrate \spine{} as a plugin into \piagent{}~\cite{earendil2026pirepo}, a popular open-source coding-agent harness with an extensible architecture~\cite{pi2026npm}, through its official extension interface~\cite{earendil2026piextensions}. Using \geminithreeeightflash{}~\cite{google2026gemini38flash} and the open-weight \qwenthreeeighttwentysevenb{}~\cite{qwen2026qwen38}, this integration improves \swemilestone{} Score by 9.5--17.5\% and the average \programbench{} test-pass rate by 4.3--8.7\%, compared with native \piagent{}.%

This paper makes the following contributions:

\begin{itemize}[leftmargin=*]
  \item \textbf{\spine{}.} We propose an interface that aligns model context with subtask nesting and completion while supporting both local execution and delegation for agents. It combines explicit subtask declarations with context construction from a persistent execution tree, namely \spinetree{}, allowing the current agent to execute nested subtasks with a structurally aligned interaction history.
  \item \textbf{System Implementation}. \modify{We integrate \spine{} with the existing mechanisms of \codex{} and \piagent{}, yielding \spinecodex{} and a loadable \spinepi{} plugin, respectively. The implementation involves approximately 17K lines of Rust code and 9K lines of TypeScript/Rust code, which are ready to use out of the box.}
  \item \textbf{\modify{Extensive Study}.} \modify{We conduct an extensive study across three long-horizon coding benchmarks, four models, and two agent harnesses. On \swemilestone{} with \gptfivesixsol{}, \spinecodex{} resolves 19 of 98 milestones, compared with 14 for native \codex{} and 17 for \recursivesubagent{}, at 32.4\% and 73.7\% lower total cost, respectively. In the \piagent{} integration, \spine{} improves \swemilestone{} Score by 9.5--17.5\% and the average \programbench{} test-pass rate by 4.3--8.7\% across the evaluated models.} %
\end{itemize}

\section{Motivating Example}
\label{sec:motivation}

\begin{figure}[htb]
  \centering
  \includegraphics[width=0.95\columnwidth]{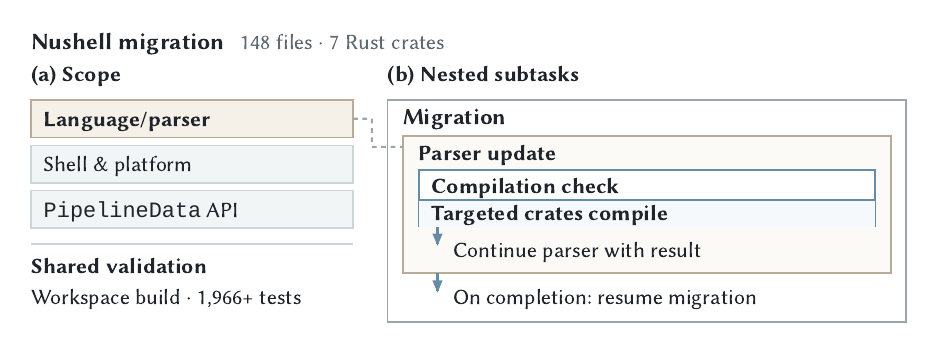}
  \caption{Nested work within the \nushell{} migration.}
  \label{fig:motivation-task}
  \Description{An illustrative decomposition of the migration contains a parser update with a nested compilation check, alongside other changes and workspace validation.}
\end{figure}

The task is a version migration of \nushell{}, i.e., a shell and scripting language, requiring coordinated changes across the repository~\cite{deng2026evoclaw}. Within a parser update, for example, the agent can run a targeted compilation check and then resume the unfinished update as in Figure~\ref{fig:motivation-task}. At this stage, subsequent model inputs require the check result alongside the ongoing parser work, whereas the check's intermediate interactions need not all remain.

\begin{figure}[htb]
  \centering
  \includegraphics[width=0.95\columnwidth]{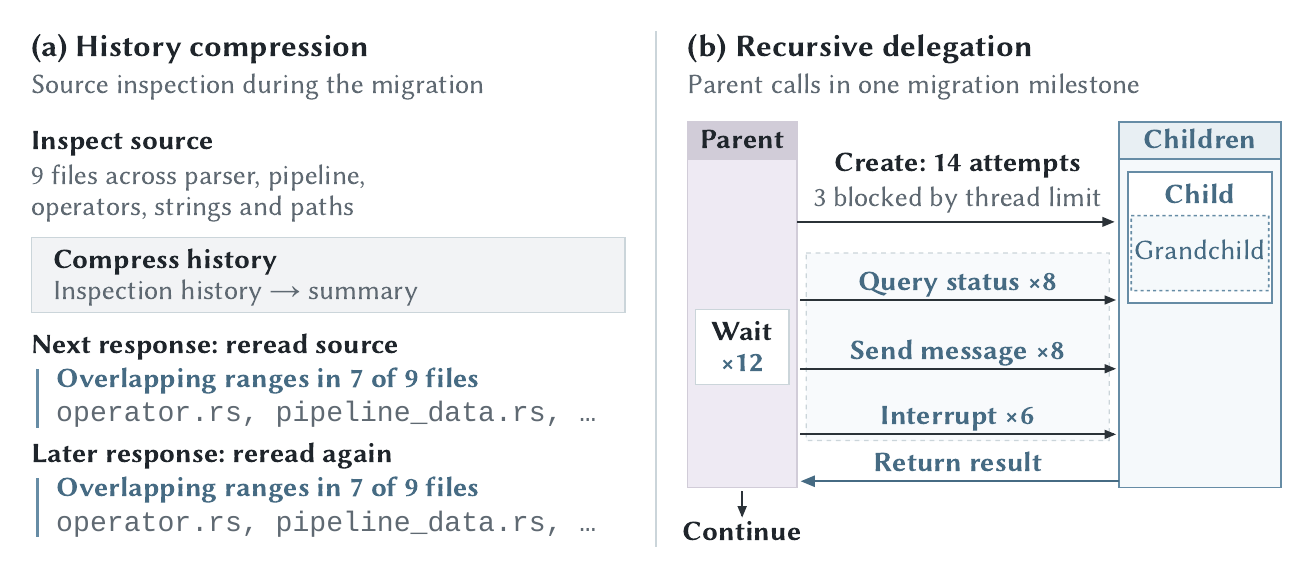}
  \caption{History compression and recursive delegation during the \nushell{} migration.}
  \label{fig:motivation-history}
  \Description{The history-compression sequence inspects nine source files, reduces an inspection slice, then rereads overlapping ranges from seven of those files in each of two later responses. The recursive-delegation panel shows nested child executions and parent calls in one migration milestone: fourteen creation attempts, three blocked by the thread limit, eight status queries, twelve wait calls, eight messages, and six interrupts.}
\end{figure}

We first consider an agent that compresses a flat interaction history without explicitly representing the nested structure of its subtasks~\cite{xiao2026agentdiet}. The reduced context must still allow the model to identify which inspections have already been completed and where implementation should resume. In Figure~\ref{fig:motivation-history}(a), the agent inspects nine source files, and the resulting inspection history is replaced by a summary. The next response rereads overlapping ranges from seven of these files, and a later response revisits seven again. In this trace, the agent revisits previously inspected source while implementation remains unfinished. These repeated inspections add tool calls and bring the same source content back into model input, consuming additional execution budget. The challenge is to reduce completed subtasks' details while keeping the context of unfinished work available.

Next, we consider an agent that delegates subtasks to child agents, which can in turn delegate further subtasks. Each child has a local context and returns a result to its parent, giving nested work separate execution boundaries. The parent must assign work, communicate requirements, check progress, and collect results. If subsequent work depends on a child's result, the parent must wait for it before proceeding. In the milestone shown in Figure~\ref{fig:motivation-history}(b), the parent issues twelve wait calls and makes fourteen attempts to create child agents, three of which are blocked by the thread limit. These handoffs and coordination steps add model and tool interactions alongside the child agents' own execution. \modify{Thus, relying on delegation merely to isolate context introduces substantial coordination overhead, particularly when the parent simply blocks to await sequential results.} %

\begin{figure}[htb]
  \centering
  \includegraphics[width=0.92\columnwidth]{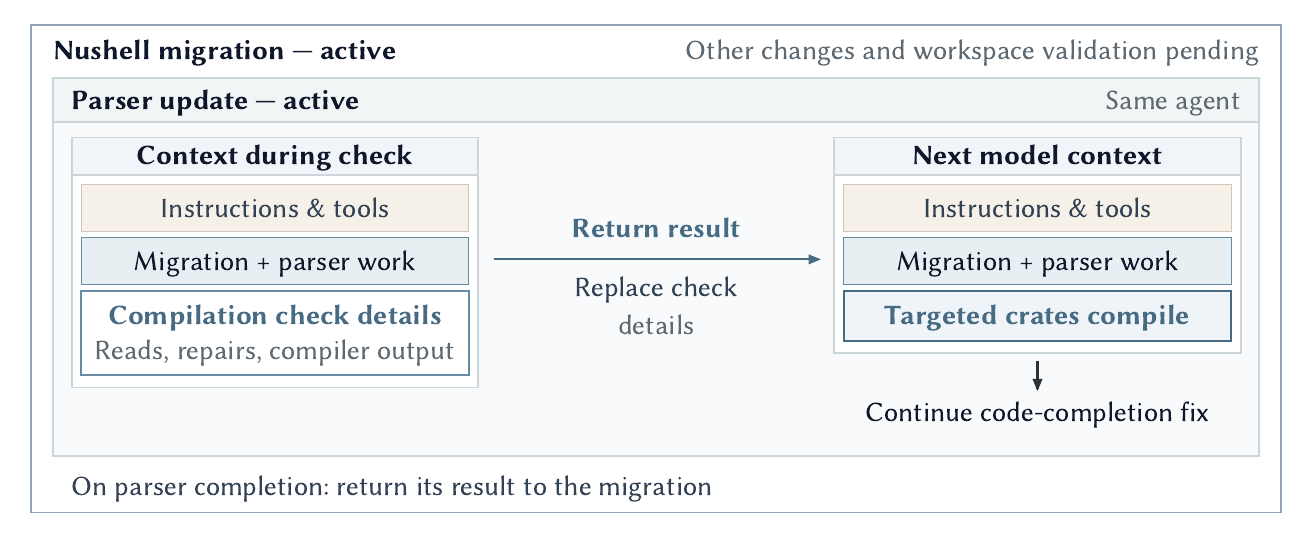}
  \caption{Returning nested results within the same agent.}
  \label{fig:motivation-ideal}
  \Description{The compilation check completes inside the parser update. Its result replaces its details in subsequent context. The same agent continues the parser update, whose later result returns to the migration with other work still pending.}
\end{figure}

Ideally, an agent could complete the nested compilation check and resume the parser update with its earlier context and the check result within its own execution, as in Figure~\ref{fig:motivation-ideal}. The agent would first mark the check's beginning, allowing the runtime to record its interactions separately from the ongoing parser work. When the check completes, the agent would return a result such as ``targeted crates compile''. The runtime would replace the check's intermediate reads, repairs, and compiler output with this result in the next model input, while retaining the context of the unfinished parser update. With the check result available, the same agent could continue the parser update \modify{for local-variable completion}. When the parser update finishes, its result would return to the migration in the same way, while other changes and workspace validation remain pending. This example motivates explicit subtask boundaries that let the current agent replace completed subtasks' details with their results while retaining information needed by unfinished parent tasks \modify{and leaving} the choice of executing agents independent of the boundary.

\section{Background and Related Work}
\label{sec:related-work}

\subsection{Long-Horizon Coding Agents and Tasks}

\begin{figure}[htb]
  \centering
  \includegraphics[width=0.85\columnwidth]{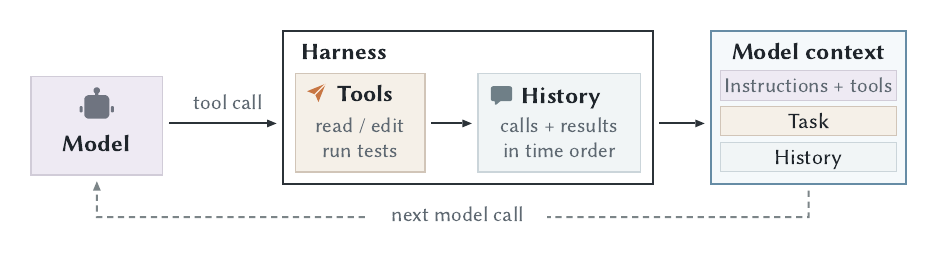}
  \caption{The coding-agent execution loop.}
  \label{fig:background-concepts}
\end{figure}

\modify{Coding-agent harnesses, e.g., \codex{}, \claudecode{}, and \piagent{}, let an agent inspect, modify, and test a repository through repeated model calls and tool use~\cite{yang2024sweagent,wang2025openhands,openai2026codexrepo,Anthropic-ClaudeCode-DynamicWorkflows,earendil2026pirepo,opencode2026repo,deepseek2026harness,cursor2026longrunning,anthropic2026managedagents}. %
As shown in Figure~\ref{fig:background-concepts}, the harness repeatedly executes tool calls, records their results, and prepares the next model input. As the loop repeats, the harness accumulates interactions and selects what context to include in each subsequent model call. Typically, the interaction history refers to the record of these interactions and model context refers to the input to one model call, including instructions, the task, and selected history.}

In long-horizon coding tasks, agents use this loop to carry out a sequence of interdependent development steps. Researchers have developed benchmarks to evaluate agents on such tasks. For example, \swemilestone{} requires agents to complete dependent milestones in a persistent repository~\cite{deng2026evoclaw}; \programbench{} requires whole-program reconstruction from documentation and an execute-only reference implementation~\cite{yang2026programbench}; and \frontierswe{} covers open-ended implementation, performance optimization, and research tasks~\cite{proximal2026frontierswe}.

To complete these tasks, an agent may perform subtasks within an ongoing subtask, each of which can span several iterations of this loop. When a subtask completes, the agent uses its result to resume the parent task, i.e., the task that directly contains it. Figure~\ref{fig:motivation-task}(b) illustrates this recursive structure. Specifically, the compilation check returns its result to the parser update, which later returns its own result to the migration. When the same agent performs these nested subtasks, the harness records their interactions in one chronological history.

This order alone does not mark where the check begins and ends or which interactions its unfinished parent still needs. A harness can compact earlier interactions to keep later inputs within the context window, while the reduced input must retain the information needed to continue unfinished work~\cite{openai2025codexmax,young2025harnesses}. With exact-prefix caching, rewriting an earlier token prevents reuse of cached computation beyond that point, while the unchanged prefix remains reusable~\cite{zheng2024sglang,dontbreakcache2026}.

\subsection{Context-Management Approaches}

Recent approaches manage model context by selecting, condensing, or retaining information across calls. For example, \agentdiet{} uses a separate reflection model to remove redundant or expired material within a sliding window of history~\cite{xiao2026agentdiet}, while \agentfold{} lets the agent summarize selected historical spans at different granular levels~\cite{ye2025agentfold}. Other approaches provide context condensation as a callable operation~\cite{liu2026contexttool}. For coding agents, \swepruner{} selects task-relevant lines from incoming file observations~\cite{wang2026swepruner}, whereas \attncompress{} uses proxy attention to select blocks from accumulated trajectories~\cite{zeng2026attncompress}.

To preserve input information needed for later work, recent approaches improve compression instructions or let the agent restore omitted context~\cite{kang2026acon,li2026sculptor}. \swemem{} lets the agent choose when, what span, and how to compress based on task progress~\cite{gao2026swemem}. \autocompact{} trains a coding agent to compact after a stage is sufficiently resolved~\cite{zhang2026autocompact}. \corvus{} keeps file observations synchronized with repository changes~\cite{zheng2026corvus}, while \ledger{} maintains execution facts and mediates repeated tool calls~\cite{wang2026ledger}.

Other approaches use subtask completion to decide when to replace detailed interactions. \hiagent{} keeps detailed interactions for the current subgoal and summarizes completed subgoals, while allowing their full trajectories to be retrieved~\cite{hu2025hiagent}.
\contextfolding{} uses \foldgrpo{} to train agents in a single-level plan--branch--return workflow to learn when to create a branch for a subtask and when to return. Specifically, a completed branch's returned result replaces its intermediate interactions in the parent context~\cite{sun2025contextfolding}.%

For approaches that trim or summarize context without explicitly maintaining work structure, the model must reconstruct subtask relations, completion status, and pending work from retained interactions. Long-running coding observations report increased reading and searching after compaction~\cite{deng2026evoclaw}, as well as difficulty recovering progress and premature completion across context windows~\cite{young2025harnesses}. These problems add execution cost and disrupt continued progress. Context management therefore needs to both preserve work state and reduce input.

\subsection{Recursive Subagent Systems}

Recursive calls give each child its own context and let it invoke further children before returning a result to its parent~\cite{yang2026recursive,schroeder2025thread,rah2026}. \rah{} applies this pattern to a complete tool-using harness, such that each child has its own context, workspace, and tools~\cite{rah2026}. \thread{} recursively spawns child execution threads with contexts derived from their parents and returns selected outputs to the parent threads~\cite{schroeder2025thread}. \modify{Current advanced coding-agent harnesses also expose this delegation pattern; for example, \codex{} can spawn child sessions from a parent thread and return their results, while \claudecode{} can invoke subagents through programmatic workflows~\cite{openai2026codexmultiagentv2,Anthropic-ClaudeCode-DynamicWorkflows}.}

\modify{However, when }subtask boundaries are established through child-agent calls, finer-grained task decomposition also requires more agent executions and communication. \agentprune{} finds that some messages contribute little to task completion in the evaluated multi-agent systems, yet are still encapsulated as the receiving agents' model inputs and consume tokens~\cite{zhang2025agentprune}. \kvcomm{} shows that when shared content follows different prefixes, ordinary prefix caching cannot directly reuse its computation, such that agents process the same content again~\cite{ye2025kvcomm}. Recursive subtask boundaries should therefore be represented independently of whether the work remains with the current agent or is delegated to another agent.

\section{\spine{} Approach}
\label{sec:method-draft}
\label{sec:spine}

\spine{} is an interface that aligns model context with subtask nesting and completion in a just-in-time manner. \modify{Specifically, it addresses the three issues illustrated in Section~\ref{sec:introduction} as follows:
(1)~the agent declares explicit subtask boundaries and completion status through \workunits{} and a persistent \spinetree{} (Section~\ref{sec:context-tree});
(2)~the interface decouples subtask granularity from executing agent choice by supporting both local inline nesting and parallel delegation (Section~\ref{sec:commands}); and
(3)~the runtime dynamically distinguishes completed subtasks from unfinished subtasks by compiling context just in time, resuming parent tasks with submitted results while preserving prefix caches (Section~\ref{sec:compilation}).}

\begin{figure*}[htbp]
  \centering
  \includegraphics[width=0.99\textwidth]{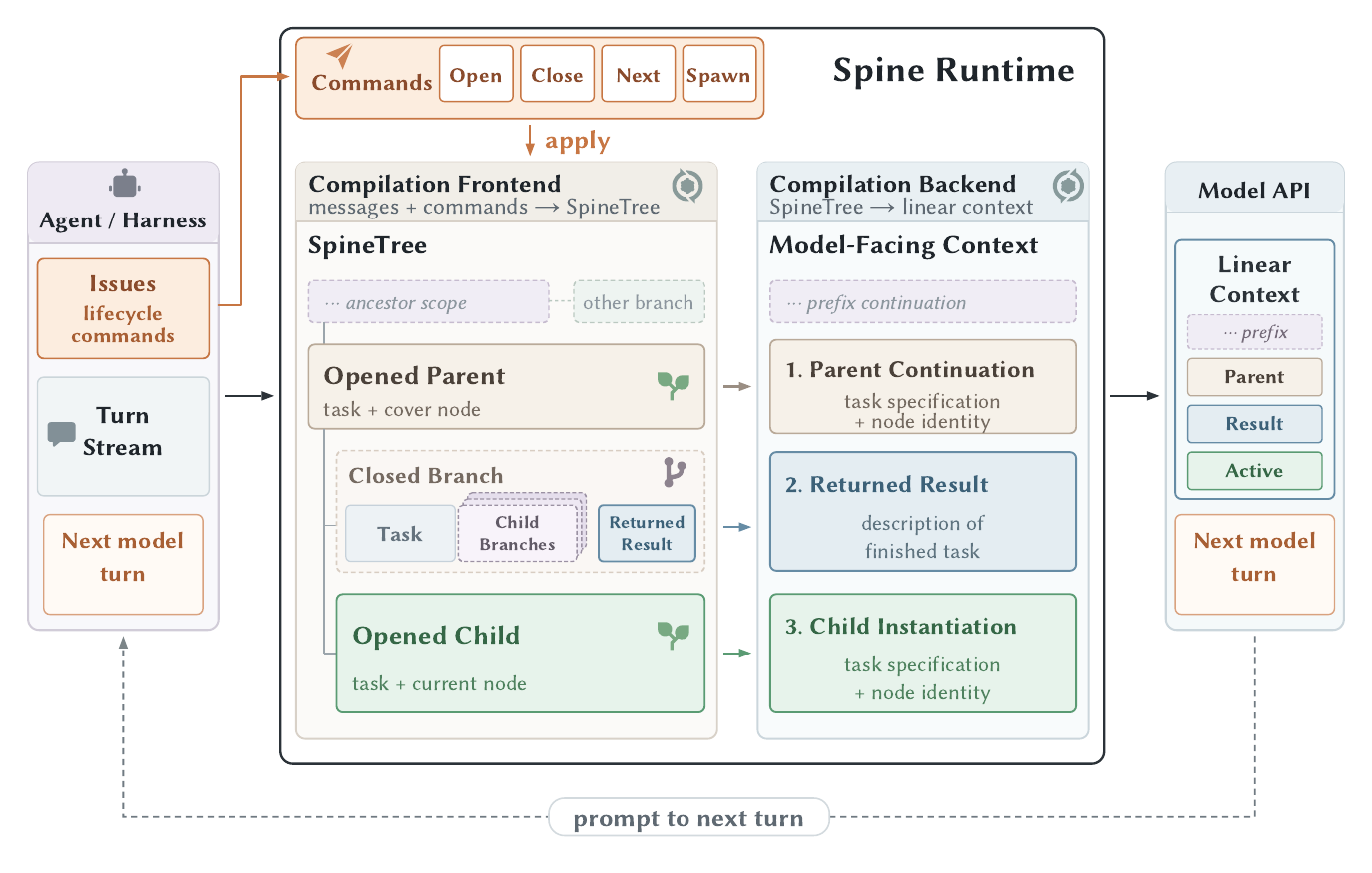}
  \caption{Overview of \spine{}.}
  \label{fig:spine-workflow}
  \Description{The Agent/Harness sends a turn stream and Open, Close,
  Next, and Spawn commands to the Compilation Frontend. Its SpineTree
  contains an open parent, a closed branch with retained task and child
  records plus a returned result, and an open child. The Compilation Backend
  maps these to parent continuation, returned result, and child instantiation,
  respectively, following the preceding context prefix. The resulting linear
  context goes to the Model API for the next turn. A dashed return path
  connects that turn to the Agent/Harness.}
\end{figure*}

Figure~\ref{fig:spine-workflow} shows how the runtime supports this
interaction. The agent issues lifecycle commands through its harness,
and the \emph{frontend} incorporates these commands and the associated
messages into \spinetree{}, preserving work nesting and interaction order.
The \emph{backend} uses the updated tree to assemble the next model input.
\modify{Specifically, as illustrated, this context retains the parent task's preceding messages and the active child's ongoing interactions, while replacing completed child interactions with their returned results.}

\subsection{Work Scopes and \spinetree{}}
\label{sec:context-tree}

\modify{%
The fundamental unit of interaction with \spine{} is a \workunit{}, i.e., an agent-declared subtask with an explicit entry, a parent task, and a result submitted on completion.
Within a scope, the current agent can inspect files, edit code, and run checks over multiple turns. Entering a nested subtask forms a child scope whose result returns to the unfinished parent. We call work performed by the current agent \emph{inline work}, whereas delegated work is performed by another agent.}

\modify{%
\spinetree{} is an ordered, persistent execution tree that serves as the structural spine of the execution. Specifically, the sequence of unmatched $\mathsf{\sjOpenName}$ markers represents the active path from the root task to the current subtask, while completed inline scopes form closed branches $b = \mathsf{\sjBranchSymbol}(d,F,\mu)$ that retain their local contents and returned results off the active path.}

Let $\mathcal{M}$ be the set of messages. Formally, a branch is defined as
\begin{equation}
  b = \mathsf{\sjBranchSymbol}(d,F,\mu),
  \qquad \mu\in\mathcal{M}^{*}.
  \label{eq:spinebranch}
\end{equation}
Its components are as follows.
\begin{description}
  \item[$d$ (Task):]
  the task description supplied when the scope is opened.
  \item[$F$ (Branch Body):]
  an ordered list $F=[Y'_1,\ldots,Y'_k]$ of this scope's local message
  nodes, completed child branches, and delegated result nodes, preserving
  their interleaving.
  \item[$\mu$ (Returned Result):]
  a message sequence $\mu$ submitted on completion to the parent task.
  Subsequent model context uses this result in place of the details retained
  in $F$.
\end{description}

While a scope remains open, the runtime must also record where its work
began and which interactions belong to it. An ordered list $T$ encodes
\spinetree{} as a whole, including messages, unfinished scopes, completed
branches, and delegated results. Each $F$ is the body of one completed
$\mathsf{\sjBranchSymbol}(d,F,\mu)$, which appears as an entry in $T$ or
in an enclosing branch body. Using $Y$ for an entry, we define
\begin{equation}
  \begin{aligned}
    T &= [Y_1,\ldots,Y_n],\\
    Y &::= \mathsf{Message}(m)\mid\mathsf{\sjOpenName}(d)
       \mid \mathsf{\sjBranchSymbol}(d,F,\mu)
       \mid\mathsf{\sjSpawnNodeSymbol}(\mu).
  \end{aligned}
  \label{eq:context-tree}
\end{equation}
Here $\mathsf{Message}(m)$ wraps a message $m\in\mathcal{M}$,
$\mathsf{\sjOpenName}(d)$ marks unfinished inline work, and
$\mathsf{\sjSpawnNodeSymbol}(\mu)$ stores a delegated result.
Node identities are implicit in this notation.

In $T$, each $Y_i$ belongs to the nearest preceding
$\mathsf{\sjOpenName}(d)$ at position $j<i$, or the implicit root if
none exists. Each entry of $F$ belongs directly to its enclosing branch.
Open markers identify the active ancestor chain: \modify{the innermost marker denotes the active subtask, or execution resides at the root task if none remain}. Note that completed branches contain
no Open markers, since child scopes must finish before their parent scope
closes. As a result, these rules preserve work ownership alongside message order.

\subsection{Interface Operations}
\label{sec:commands}

\modify{The interface provides four commands for organizing work during task execution, decoupling subtask nesting from executor choice. The agent uses \openop{} to enter a child scope locally and \closeop{} to complete it with a result, allowing the same agent to execute nested subtasks inline without multi-agent overhead. Once the child completes, the parent resumes with that result in its model context. \nextop{} combines completion with entry into a sibling scope, while \spawnop{} delegates independent subtasks to other agents in parallel and collects their results.} Ordinary model and tool interactions continue within the current scope between these commands.

To apply these operations to \spinetree{}, the runtime translates messages
and command events into \modify{\emph{terminals}---discrete input tokens representing messages, boundaries, and results}. The frontend
(Section~\ref{sec:frontend}) consumes the resulting stream in order.

A turn comprises one model interaction and its associated harness processing.
\modify{Assume $P$ contains pending incoming messages (such as user prompts or delegated returns) awaiting processing at turn start}, and $O$ contains the turn's local messages, including tool calls and results.
For $P=[p_1,\ldots,p_k]$ and $O=[o_1,\ldots,o_\ell]$, mapping each
message to a parser terminal gives
\begin{equation}
  \begin{aligned}
    \widehat P&=[\mathsf{Message}(p_1),\ldots,\mathsf{Message}(p_k)],\\
    \widehat O&=[\mathsf{Message}(o_1),\ldots,\mathsf{Message}(o_\ell)].
  \end{aligned}
  \label{eq:terminal-encoding}
\end{equation}
The hats denote the same messages represented as parser terminals, where their
content and order are unchanged. Each command produces a terminal stream
$\omega$, consumed by the frontend in order. Here, ``$\cdot$'' denotes
concatenation, and delegated results are appended as separate result nodes.
Table~\ref{tab:lifecycle-commands} summarizes the corresponding state
transitions and streams. An \openop{} enters a child\modify{~scope} under the current scope,
or under the root when no scope is open. A \closeop{} retains the child\modify{~scope}'s body
$F$ in \spinetree{} while replacing it with result $\mu$ in the parent's next
model input; \closeop{} and \nextop{} require an open non-root scope.
\begin{table}[t]
  \caption{\footnotesize Lifecycle commands and terminal streams. $\widehat P$
  and $\widehat O$ encode preceding and current-turn messages; ``$\cdot$''
  denotes list concatenation; delegated results follow task order.}
  \label{tab:lifecycle-commands}
  \centering
  \footnotesize
  \setlength{\tabcolsep}{2pt}
  \renewcommand{\arraystretch}{1.15}
  \begin{tabular}{@{}p{0.16\linewidth}>{\centering\arraybackslash}p{0.49\linewidth}>{\raggedleft\arraybackslash}p{0.27\linewidth}@{}}
    \toprule
    Command/event & Terminal stream $\omega$ & Work semantics \\
    \midrule
    No command &
    $\widehat P\cdot\widehat O$ &
    Continue current work. \\[2pt]
    \openop{}$(d)$ &
    $\widehat P\cdot[\mathsf{\sjOpenName}(d)]\cdot\widehat O$ &
    Enter inline child. \\[2pt]
    \closeop{}$(\mu)$ &
    $\widehat P\cdot[\mathsf{\sjCloseName}(\mu)]\cdot\widehat O$ &
    Complete child; return result. \\[2pt]
    \nextop{}$(\mu,d)$ &
    $\widehat P\cdot[\mathsf{\sjCloseName}(\mu),\mathsf{\sjOpenName}(d)]
      \cdot\widehat O$ &
    Close child; open sibling. \\[2pt]
    \spawnop{}$(tasks)$ &
    $\widehat P\cdot\widehat O\cdot
      [\mathsf{\sjSpawnNodeSymbol}(\mu_i)]_{i=0}^{k-1}$ &
    Delegate tasks; collect results. \\
    \bottomrule
  \end{tabular}
\end{table}

\subsection{\modify{Just-In-Time Context Compilation}}
\label{sec:compilation}

\begin{figure}[htbp]
  \centering
  \includegraphics[width=0.88\columnwidth]{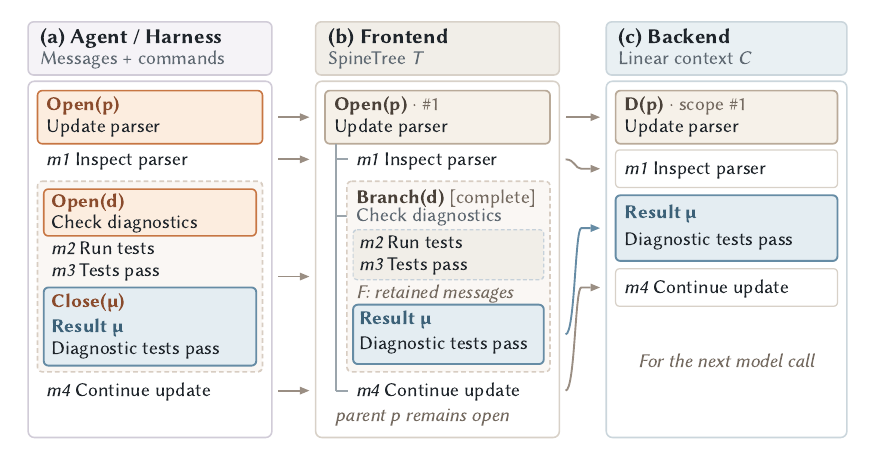}
  \caption{Context compilation for an illustrative diagnostic check.}
  \label{fig:backend-projection}
  \Description{Three panels follow a parser update. The agent opens the
  parent work, records message m1, opens a diagnostic check, records a test
  call m2 and output m3, and closes the check with result mu before continuing
  with parent message m4. SpineTree retains m2 and m3 in the completed
  branch body F alongside mu. The next linear input contains the parent's
  task and identity D(p), m1, mu, and m4 in order.}
\end{figure}

\modify{Context compilation implements the interface operations by dynamically assembling model context from the persistent work records \emph{just in time} before each model call.}
Figure~\ref{fig:backend-projection} illustrates a diagnostic check within
an unfinished parser update. Specifically, the agent declares the check's boundaries
through \openop{} and \closeop{}, submitting ``Diagnostic tests pass'' as
its result $\mu$. These boundaries assign the test call and output
($m_2,m_3$) to the check. The frontend stores them in the completed branch's
body $F$. When the parent continues, the backend carries $\mu$ into the
next input alongside the parent's task and identity, $D(p)$, and its
messages ($m_1,m_4$). Note that the test details remain in \spinetree{}.

\subsubsection{Compilation frontend}
\label{sec:frontend}

In turn $t$, with pending messages $P_t$, local messages $O_t$, and command
event $x_t$, the frontend incrementally incorporates the terminal stream
$\omega_t$ into the existing tree $T_t$:
\begin{equation}
  \begin{aligned}
    \omega_t &= \operatorname{terminals}(P_t,O_t,x_t),\\
    T_{t+1} &= \operatorname{frontend}(T_t,\omega_t).
  \end{aligned}
  \label{eq:frontend}
\end{equation}
\modify{At its core, scope lifecycle follows a balanced-bracket structure. Specifically, an $\mathsf{\sjOpenName}$ marker initiates a scope, intermediate messages accumulate in its body, and a matching $\mathsf{\sjCloseName}$ marker seals the scope into a completed branch. Formally, the following LR(0) grammar~\cite{aho1974lrparsing} specifies this process without lookahead, constructing branches deterministically as soon as their $\mathsf{\sjCloseName}$ terminals arrive.}
Note that we assume nonempty streams and branch bodies. With payloads omitted,
the grammar is
\begin{equation}
  \begin{array}{rcl}
    \mathsf{Start} &\to& \mathsf{Nodes}\ \mathsf{End}\\
    \mathsf{Nodes} &\to& \mathsf{Node}\mid\mathsf{Nodes}\ \mathsf{Node}\\
    \mathsf{Node} &\to& \mathsf{Message}\mid\mathsf{\sjReducedNodeSymbol}\\
    \mathsf{\sjReducedNodeSymbol}
      &\to& \mathsf{\sjOpenName}\ \mathsf{Nodes}\ \mathsf{\sjCloseName}
      \mid\mathsf{\sjSpawnNodeSymbol}.
  \end{array}
  \label{eq:grammar}
\end{equation}
The runtime incrementally parses prefixes of valid streams.
$\mathsf{End}$ is a logical end marker which is never emitted during execution.
The \parsestack{} retains parser states and semantic values across turns.

Operationally, the frontend shifts each terminal from $\omega$ left to
right onto the \parsestack{}. After each shift, it repeatedly applies
reductions selected by the current LR(0) state, replacing the matching
stack suffix with the production's left-hand-side symbol.
Reducing $\mathsf{\sjOpenName}\ \mathsf{Nodes}\ \mathsf{\sjCloseName}$ to
$\mathsf{\sjReducedNodeSymbol}$ constructs the innermost
$\mathsf{\sjBranchSymbol}(d,F,\mu)$ from the opening task, ordered body,
and closing result. The delegated rule retains
$\mathsf{\sjSpawnNodeSymbol}(\mu)$, while $\mathsf{Nodes}$ reductions preserve node
order. After all available reductions, flattening the stack's semantic
node lists in place yields $T$  with unmatched Open markers retained
(Section~\ref{sec:context-tree}). The backend can therefore project the tree
while open work continues across turns.

\subsubsection{Compilation backend}
\label{sec:projection}

The backend assembles model input from the updated work state. It carries
messages and open task descriptors forward, while using each completed
child's result in place of its detailed interactions.
For $T_{t+1}=[Y_1,\ldots,Y_n]$, this projection is
\begin{equation}
  \begin{aligned}
    C_{t+1}
      &=\operatorname{backend}(T_{t+1})
        =h(Y_1)\cdot\ldots\cdot h(Y_n),\\
    h(Y)&=
    \begin{cases}
      [m], & Y=\mathsf{Message}(m),\\
      D(d), & Y=\mathsf{\sjOpenName}(d),\\
      \mu, & Y\in\{\mathsf{\sjBranchSymbol}(d,F,\mu),
                    \mathsf{\sjSpawnNodeSymbol}(\mu)\}.
    \end{cases}
  \end{aligned}
  \label{eq:projection}
\end{equation}

Here $D(d)\in\mathcal{M}^{*}$ renders an Open node's task and node identity
as messages. \modify{Crucially, a completed inline branch contributes solely its returned result $\mu$ without expanding its body $F$, shielding subsequent turns from voluminous intermediate interactions while safely preserving them in \spinetree{}.} A delegated node
contributes its result, while the delegated agent manages the local interactions.

\noindent\textbf{Prefix Cache Preservation.}
A delegated child's input extends its parent's context with a task-specific
suffix $\Delta_{\mathrm{child}}$. \modify{Completing an inline scope} affects only the suffix
of $T$ beginning at the active Open marker. Its projection $B$, including
the task description, is replaced by $\mu$, preserving the preceding
prefix $Q$ of the model context:
\begin{equation}
  \begin{aligned}
    C_{\mathrm{child}}
      &= C_{\mathrm{parent}}\cdot\Delta_{\mathrm{child}},\\
    Q\cdot B &\longrightarrow Q\cdot\mu.
  \end{aligned}
  \label{eq:prefix-cache}
\end{equation}
The second line isolates branch replacement, where any new sibling task and the
current turn's messages are then appended as in Section~\ref{sec:commands}.
With identical token serialization and compatible cache settings, the
unchanged prefixes remain eligible for KV-cache reuse, although actual reuse depends
on cache availability.
\modify{Specifically, inference engines can reuse the precomputed key-value states for $Q$, avoiding redundant attention computation across turns.}

\section{Evaluation}

\label{sec:evaluation}

\subsection{Benchmarks and Evaluation Metrics}
\label{sec:eval-bench-metrics}

We evaluate \spine{} on three widely-adopted benchmarks covering continuous repository evolution, whole-program reconstruction, and open-ended engineering. Table~\ref{tab:benchmarks} summarizes the tasks evaluated and the metrics used for each benchmark.

\begin{table}[!h]
  \caption{Evaluation benchmarks and metrics.}
  \label{tab:benchmarks}
  \centering
  \small
  \renewcommand{\arraystretch}{1.08}
  \setlength{\tabcolsep}{4pt}
  \begin{tabular*}{0.8\columnwidth}{@{}l@{\extracolsep{\fill}}ccr@{}}
    \toprule
    \textbf{Benchmark} & \textbf{Scope} & \textbf{Metrics} & \textbf{Target} \\
    \midrule
    \swemilestone{}~\cite{deng2026evoclaw}
      & \makecell[c]{7 tasks\\(98 milestones total)}
      & \makecell[c]{Score (\%)\\\# Resolved}
      & \makecell[r]{Continuous\\repository evolution} \\
    \addlinespace[3pt]
    \programbench{}~\cite{yang2026programbench}
      & \makecell[c]{50 tasks randomly\\sampled from 200}
      & \makecell[c]{\# Resolved / \# Almost\\Avg. pass (\%)}
      & \makecell[r]{Whole-program\\reconstruction} \\
    \addlinespace[3pt]
    \frontierswe{}~\cite{proximal2026frontierswe}
      & \makecell[c]{10 tasks\\5 trials/task}
      & \makecell[c]{Mean@5\\Best@5}
      & \makecell[r]{Open-ended\\engineering} \\
    \bottomrule
  \end{tabular*}
\end{table}

\begin{itemize}[leftmargin=*]
\item\textbf{\swemilestone{}~\cite{deng2026evoclaw}} requires agents to complete dependent milestones in a persistent repository (e.g., upgrading go-zero's go-redis dependency from v8 to v9~\cite{swemilestone2026gozero}) while preserving existing behaviors.

\item\textbf{\programbench{}~\cite{yang2026programbench}} requires agents to reconstruct programs ranging from command-line utilities to \ffmpeg{} and \sqlite{}~\cite{programbench2026ffmpeg,programbench2026sqlite} using documentation and an execute-only reference implementation. Agents produce source code and build files, which are evaluated through hidden tests that check whether the resulting program reproduces the reference behavior.

\item\textbf{\frontierswe{}\allowbreak~\cite{proximal2026frontierswe}} covers open-ended engineering tasks, such as building a \postgresql{}-compatible server backed by \sqlite{}, implementing \git{} in \zig{}, and optimizing \texttt{\libswscale{}} ~\cite{frontierswe2026postgresqlsqlite,frontierswe2026gittozig,frontierswe2026libswscale}. Continuous scores measure partial progress toward each task's objective.

\end{itemize}

\modify{Across all three benchmarks, we report their official evaluation metrics.} For \swemilestone{}, %
\modify{we report its official metric \textbf{Score (\%)}~\cite{deng2026evoclaw}, which evaluates milestone progress as the harmonic mean of test recall (new functionality) and precision (regression prevention), and the number of milestones passing all required tests (\textbf{\# Resolved}).} For \programbench{}~\cite{yang2026programbench}, we report the numbers of tasks passing all tests (\textbf{\# Resolved}) or at least 95\% of tests (\textbf{\# Almost}, including fully resolved tasks), together with the average test-pass rate across tasks (\textbf{Avg. pass (\%)}). For \frontierswe{}~\cite{proximal2026frontierswe}, we report the mean (\textbf{Mean@5}) and maximum (\textbf{Best@5}) scores over five trials per task.

We also report monetary cost (\textbf{Cost (\$)}), total token usage (\textbf{Total tokens}), and average input tokens per main-model request (\textbf{Avg. input tokens/req.}).

\subsection{Evaluation Harness Systems and Models}
\label{sec:eval-systems}

\modify{To evaluate \spine{} across different types of agent harnesses, we select two popular open-source systems: \codex{}~\cite{openai2026codexrepo}, which features a native session lifecycle, and \piagent{}~\cite{earendil2026pirepo}, which provides an extensible plugin architecture.}
\modify{We implement \spine{} on them as \spinecodex{} and \spinepi{}, respectively, both of which are ready to use out of the box with their original harnesses, \basecodex{} and \basepi{}, as baseline controls.}

\begin{itemize}[leftmargin=*]
\item\textbf{\codex{} / \spinecodex{}.} We select \codex{}, whose repository has over 127,000 GitHub stars~\cite{openai2026codexrepo} and whose broader ecosystem has over five million weekly active users~\cite{openai2026codexknowledgework}, as the common harness for comparing context-management approaches in RQ1. \spinecodex{} connects to \codex{}'s native session lifecycle to support work scopes, forked child sessions, and execution resumption\modify{, with approximately 17K lines of Rust code}.

\item\textbf{\piagent{} / \spinepi{}.} We select \piagent{}, whose repository has over 111,000 GitHub stars~\cite{earendil2026pirepo} and whose coding-agent package received over 4.5 million npm downloads in one week~\cite{pi2026npm}, to evaluate \spine{}'s \modify{generalizability} in RQ2. \spinepi{} is loaded as a plugin through \piagent{}'s extension interface, registering \spine{} tools and providing model context~\cite{earendil2026piextensions}\modify{, with approximately 9K lines of TypeScript/Rust code}.
\end{itemize}

\paragraph{Baselines.}
We select four baselines covering compaction, trajectory reduction, context folding, and recursive delegation. All approaches run within \codex{}.

\begin{itemize}[leftmargin=*]
\item\textbf{\basecodex{}}. We use \codex{}'s native context compaction mechanism~\cite{openai2026codexrepo,openai2025codexmax}.%

\item\textbf{\agentdiet{}.} \agentdiet{} uses reflection over a sliding window of the trajectory to identify and remove redundant information~\cite{xiao2026agentdiet}.

\item\textbf{\contextfolding{}.} \contextfolding{} lets the agent branch into a subtask and return a summary to the parent context. We implement its folding mechanism with the shared model, without \foldgrpo{} training~\cite{sun2025contextfolding}.

\item\textbf{\recursivesubagent{}.} \modify{Following recursive delegation frameworks~\cite{rah2026,schroeder2025thread}, we instantiate this baseline by adapting \codex{}'s native subagent mechanism~\cite{openai2026codexmultiagentv2}, where agents recursively delegate subtasks to child sessions and incorporate their returned results.}

\end{itemize}

\paragraph{Models.}
We select models spanning frontier capabilities, cost-efficient inference, and open-weight deployment. Specifically, we use \gptfivesixsol{} and \grokfoursix{} with \codex{} for RQ1~\cite{openai2026gpt56sol,xai2026grok46}, and \geminithreeeightflash{} and \qwenthreeeighttwentysevenb{} with \piagent{} for RQ2~\cite{google2026gemini38flash,qwen2026qwen38}; we access the first three models through their providers' official APIs and run \qwenthreeeighttwentysevenb{} locally with \vllm{}~\cite{kwon2023efficient} on \nvidia{} \aOneHundred{} GPUs. %

\subsection{Evaluation Setup}
\label{sec:eval-setup}

We run each comparison in isolated containers with the same model, harness, task subset, and budget across the approaches being compared. \modify{Unless explicitly noted otherwise, our default benchmark is \swemilestone{} and our default model is \gptfivesixsol{}.}

For \swemilestone{}, we follow the official protocol and resume each run until all milestones have been submitted for evaluation~\cite{swemilestone2026protocol}. Given the cost of evaluating all 200 tasks, we randomly sample 50 tasks from \programbench{} and retain its six-hour and 1{,}000-step limits~\cite{yang2026programbench}. For \frontierswe{}, we evaluate a fixed subset of ten available V1 tasks with five trials per task and a 20-hour budget per trial. Seven of its 17 tasks~\cite{proximal2026frontierswe} are unavailable in our environment, including \texttt{optimizer-design}, which requires an \hOneHundred{} GPU~\cite{frontierswe2026optimizer}, and \texttt{frogsgame-rl}, which requires the external \tinkerapi{} and network access~\cite{frontierswe2026frogsgame}. \modify{We evaluate \frontierswe{} in RQ1 to cover open-ended engineering tasks, while RQ2 evaluates harness generalizability on \swemilestone{} and \programbench{}.}
The source code of \spinecodex{} and \spinepi{} is available on GitHub~\cite{spinecodexrepo,spinepirepo}.

\subsection{Research Questions}
We evaluate the effectiveness, \modify{generalizability}, and runtime behavior of \spine{} through three research questions:
\begin{itemize}[leftmargin=*]
  \item \textbf{RQ1: How effective is \spine{} on the \codex{} harness for long-horizon software engineering tasks?}
  To answer this RQ, we compare \basecodex{} and \spinecodex{} across three benchmarks, and then compare \spinecodex{} with other context-management approaches on \swemilestone{}.
  \item \textbf{RQ2: Does \spine{} remain effective when integrated into another agent harness?}
  To answer this RQ, we compare \basepi{} and \spinepi{} on \swemilestone{} and \programbench{} with \geminithreeeightflash{} and \qwenthreeeighttwentysevenb{}.
  \item \textbf{RQ3: How do agents use \spine{} to organize work, and how does this affect the context carried forward?}
  We analyze scope use and input changes, revisit the motivating example through an execution trace, and conduct ablation studies on \modify{subtask} nesting and boundary maintenance.
\end{itemize}

\subsection{Result Analysis}
\label{sec:eval-results}

\subsubsection{RQ1: Effectiveness of \spine{}}
\label{sec:eval-rq1}

\ifdefined\resultmodelwidth\else
  \newlength{\resultmodelwidth}
  \newlength{\resultpanelwidth}
\fi
\begingroup
\footnotesize
\setlength{\tabcolsep}{2.5pt}
\settowidth{\resultmodelwidth}{\qwenthreeeighttwentysevenb}
\sbox0{%
  \begin{tabular}{@{}p{\resultmodelwidth}lrrrrrr@{}}
    \textbf{Model} & \textbf{Agent} & \textbf{Resolved} & \textbf{Almost}
      & \makecell[r]{\textbf{Avg.}\\\textbf{pass (\%)}}
      & \makecell[r]{\textbf{Total}\\\textbf{tokens}}
      & \makecell[r]{\textbf{Avg. input}\\\textbf{tokens/req.}}
      & \textbf{Cost (\$)} \\
    \qwenthreeeighttwentysevenb & \textbf{\spinecodex{}} & 00/50 & 00/50 & 00.0 & 0.0M & 0.0K & 0 \\
  \end{tabular}%
}
\setlength{\resultpanelwidth}{\textwidth}
\global\resultmodelwidth=\resultmodelwidth
\global\resultpanelwidth=\resultpanelwidth
\endgroup

\begin{table*}[t]
  \caption{Performance of \spinecodex{} and \basecodex{} on \swemilestone{} and \programbench{}.}
  \label{tab:rq1-results}
  \centering
  \footnotesize
  \setlength{\tabcolsep}{2.5pt}
  \renewcommand{\arraystretch}{1.08}
  \newcommand{\bc}{\cellcolor{table_shallow_blue}}
  \newcommand{\tokhead}{\makecell[r]{\textbf{Total}\\\textbf{tokens}}}
  \newcommand{\avghead}{\makecell[r]{\textbf{Avg. input}\\\textbf{tokens/req.}}}
  \newcommand{\costhead}{\textbf{Cost (\$)}}

  \begin{tabularx}{\resultpanelwidth}{@{}p{\resultmodelwidth}Xrrrrr@{}}
    \toprule
    \multicolumn{7}{@{}l}{\textbf{\swemilestone{}}} \\
    \midrule
    \textbf{Model} & \textbf{Agent} & \textbf{Score (\%)} & \textbf{\# Resolved} & \tokhead & \avghead & \costhead \\
    \midrule
    \multirow{2}{\resultmodelwidth}{\gptfivesixsol}
      & \bc \basecodex{} & \bc 38.1 & \bc 14/98 & \bc 1,072.4M & \bc 131.2K & \bc 681.22 \\
      & \textbf{\spinecodex{}} & \textbf{42.6} & \textbf{19/98} & \textbf{640.5M} & \textbf{72.3K} & \textbf{460.77} \\
    \midrule
    \multirow{2}{\resultmodelwidth}{\grokfoursix}
      & \bc \basecodex{} & \bc 35.5 & \bc 12/98 & \bc 713.8M & \bc 120.0K & \bc 505.57 \\
      & \textbf{\spinecodex{}} & \textbf{40.4} & \textbf{14/98} & \textbf{444.2M} & \textbf{60.3K} & \textbf{314.20} \\
    \bottomrule
  \end{tabularx}

  \vspace{6pt}
  \begin{tabularx}{\resultpanelwidth}{@{}p{\resultmodelwidth}Xrrrrrr@{}}
    \toprule
    \multicolumn{8}{@{}l}{\textbf{\programbench{}}} \\
    \midrule
    \textbf{Model} & \textbf{Agent} & \textbf{\# Resolved} & \textbf{\# Almost}
      & \makecell[r]{\textbf{Avg.}\\\textbf{pass (\%)}} & \tokhead & \avghead & \costhead \\
    \midrule
    \multirow{2}{\resultmodelwidth}{\gptfivesixsol}
      & \bc \basecodex{} & \bc 0/50 & \bc 2/50 & \bc 62.55 & \bc 228.44M & \bc 74.71K & \bc 210.21 \\
      & \textbf{\spinecodex{}} & \textbf{2/50} & \textbf{7/50} & \textbf{73.35} & \textbf{264.32M} & \textbf{67.29K} & \textbf{221.56} \\
    \midrule
    \multirow{2}{\resultmodelwidth}{\grokfoursix}
      & \bc \basecodex{} & \bc 1/50 & \bc 5/50 & \bc 64.58 & \bc 223.53M & \bc 132.75K & \bc 189.73 \\
      & \textbf{\spinecodex{}} & \textbf{1/50} & \textbf{8/50} & \textbf{71.15} & \textbf{199.02M} & \textbf{119.16K} & \textbf{164.58} \\
    \bottomrule
  \end{tabularx}

\end{table*}

\begin{table*}[t]
  \caption{Per-task performance of \spinecodex{} and \basecodex{} on \frontierswe{}.}
  \label{tab:rq1-frontierswe}
  \centering
  \newcommand{\besttint}{\cellcolor{table_shallow_blue}}
  \footnotesize
  \setlength{\tabcolsep}{2.5pt}
  \renewcommand{\arraystretch}{1.08}
  \begin{tabularx}{\resultpanelwidth}{@{}>{\raggedright\arraybackslash}X*{8}{r}@{}}
    \toprule
    & \multicolumn{4}{c}{\textbf{\gptfivesixsol}} & \multicolumn{4}{c}{\textbf{\grokfoursix}} \\
    \cmidrule(lr){2-5}\cmidrule(lr){6-9}
    \textbf{Task}
      & \multicolumn{2}{c}{\textbf{\basecodex{}}}
      & \multicolumn{2}{c}{\textbf{\spinecodex{}}}
      & \multicolumn{2}{c}{\textbf{\basecodex{}}}
      & \multicolumn{2}{c}{\textbf{\spinecodex{}}} \\
    \cmidrule(lr){2-3}\cmidrule(lr){4-5}\cmidrule(lr){6-7}\cmidrule(lr){8-9}
    & \textbf{Mean@5} & \besttint \textbf{Best@5}
      & \textbf{Mean@5} & \besttint \textbf{Best@5}
      & \textbf{Mean@5} & \besttint \textbf{Best@5}
      & \textbf{Mean@5} & \besttint \textbf{Best@5} \\
    \midrule
    cranelift-\allowbreak codegen-\allowbreak opt & \textbf{0} & \besttint \textbf{0} & \textbf{0} & \besttint \textbf{0} & \textbf{0.0164} & \besttint \textbf{0.0822} & 0 & \besttint 0 \\
    dart-\allowbreak style-\allowbreak haskell & 0.1533 & \besttint 0.4029 & \textbf{0.4019} & \besttint \textbf{0.8136} & 0.0876 & \besttint 0.1595 & \textbf{0.4074} & \besttint \textbf{0.8505} \\
    dependent-\allowbreak type-\allowbreak checker & 1.8420 & \besttint 3.6100 & \textbf{2.4750} & \besttint \textbf{3.7400} & \textbf{0} & \besttint \textbf{0} & \textbf{0} & \besttint \textbf{0} \\
    ffmpeg-\allowbreak swscale-\allowbreak rewrite & 0.5472 & \besttint 1.6400 & \textbf{0.9488} & \besttint \textbf{1.7200} & 0.2140 & \besttint 0.6280 & \textbf{0.3910} & \besttint \textbf{0.6550} \\
    git-\allowbreak to-\allowbreak zig & 0.1149 & \besttint 0.1845 & \textbf{0.1496} & \besttint \textbf{0.2174} & 0.1085 & \besttint 0.2030 & \textbf{0.1580} & \besttint \textbf{0.2180} \\
    libexpat-\allowbreak to-\allowbreak x86asm & 0 & \besttint 0 & \textbf{0.1707} & \besttint \textbf{0.5444} & 0.0356 & \besttint 0.1781 & \textbf{0.6126} & \besttint \textbf{0.9703} \\
    lua-\allowbreak native-\allowbreak compiler & 0.5934 & \besttint \textbf{1.0000} & \textbf{0.7889} & \besttint 0.9889 & 0.6343 & \besttint 0.9006 & \textbf{0.9801} & \besttint \textbf{0.9834} \\
    postgres-\allowbreak sqlite-\allowbreak wire-\allowbreak adapter & 0 & \besttint 0 & \textbf{0.0314} & \besttint \textbf{0.1568} & \textbf{0.0595} & \besttint \textbf{0.1556} & 0.0138 & \besttint 0.0691 \\
    pyright-\allowbreak type-\allowbreak checking-\allowbreak optimization & 0.8771 & \besttint 1.6420 & \textbf{1.1263} & \besttint \textbf{1.7100} & 0.9060 & \besttint 1.1913 & \textbf{1.2808} & \besttint \textbf{1.4206} \\
    revideo-\allowbreak perf-\allowbreak opt & \textbf{0.2438} & \besttint \textbf{1.2189} & 0 & \besttint 0 & \textbf{0} & \besttint \textbf{0} & \textbf{0} & \besttint \textbf{0} \\
    \midrule
    \multicolumn{1}{l}{\textbf{Total cost (\$)}}
      & \multicolumn{2}{c}{859.78} & \multicolumn{2}{c}{\textbf{802.45}}
      & \multicolumn{2}{c}{955.30} & \multicolumn{2}{c}{\textbf{771.99}} \\
    \bottomrule
  \end{tabularx}
\end{table*}

\spinecodex{} resolves more milestones than \basecodex{} on \swemilestone{} with both evaluated models (Table~\ref{tab:rq1-results}), increasing the number from 14 to 19 with \gptfivesixsol{} and from 12 to 14 with \grokfoursix{}. The corresponding Score improves by 11.8\% with \gptfivesixsol{} and 13.8\% with \grokfoursix{}, relative to \basecodex{}.

\spinecodex{} also improves whole-program reconstruction on \programbench{} (Table~\ref{tab:rq1-results}), increasing the average test-pass rate relative to \basecodex{} by 17.3\% with \gptfivesixsol{} and 10.2\% with \grokfoursix{}. With \gptfivesixsol{}, it fully resolves two of the 50 tasks, compared with none for \basecodex{}.

On \frontierswe{}, \spinecodex{} improves Mean@5 and Best@5 over \basecodex{} on eight and seven of the ten tasks, respectively, with \gptfivesixsol{}, and improves both metrics on six tasks with \grokfoursix{} (Table~\ref{tab:rq1-frontierswe}). For example, on \texttt{libexpat-to-x86asm}, its Mean@5 and Best@5 with \gptfivesixsol{} increase from zero to 0.1707 and 0.5444, respectively.

\spinecodex{} costs less than \basecodex{} in five of the six benchmark--model settings (Tables~\ref{tab:rq1-results} and~\ref{tab:rq1-frontierswe}). The exception is \programbench{} with \gptfivesixsol{}, where the higher average test-pass rate is accompanied by a 5.4\% increase in cost.

\findingbox{\textbf{Finding 1:} \modify{\spine{} enhances \codex{}'s effectiveness across repository evolution, whole-program reconstruction, and open-ended engineering tasks, while reducing monetary cost by 6.7--37.9\% in five of the six evaluated benchmark--model settings.}}

\modify{Due to the substantial computational cost of evaluating full benchmark suites, we compare \spinecodex{} with four context-management baselines under our default setting (\swemilestone{} with \gptfivesixsol{}, Table~\ref{tab:rq1-context-tools}). Among all five approaches,} \spinecodex{} achieves the highest Score, resolves the most milestones, and uses the fewest total tokens at the lowest cost (Table~\ref{tab:rq1-context-tools})\modify{.}

\begin{table*}[t]
  \caption{Comparison with context-management approaches on \swemilestone{} using \gptfivesixsol{}.}
  \label{tab:rq1-context-tools}
  \centering
  \footnotesize
  \setlength{\tabcolsep}{2.5pt}
  \renewcommand{\arraystretch}{1.08}
  \newcommand{\bc}{\cellcolor{table_shallow_blue}}
  \begin{tabularx}{\textwidth}{@{}Xrrrrr@{}}
    \toprule
    \textbf{Agent} & \textbf{Score (\%)} & \textbf{\# Resolved}
      & \makecell[r]{\textbf{Total}\\\textbf{tokens}}
      & \makecell[r]{\textbf{Avg. input}\\\textbf{tokens/req.}}
      & \textbf{Cost (\$)} \\
    \midrule
    \bc \basecodex{} & \bc 38.1 & \bc 14/98 & \bc 1,072.4M & \bc 131.2K & \bc 681.22 \\
    \agentdiet{} & 36.5 & 12/98 & 959.2M + 76.0M\textsuperscript{*} & 92.8K & 1,057.08 + 41.44\textsuperscript{*} \\
    \bc \contextfolding{} & \bc 37.1 & \bc 13/98 & \bc 765.2M & \bc 87.3K & \bc 596.25 \\
    \recursivesubagent{} & 42.4 & 17/98 & 2,139.9M & 65.3K & 1,749.73 \\
    \bc \textbf{\spinecodex{}} & \bc \textbf{42.6} & \bc \textbf{19/98} & \bc \textbf{640.5M} & \bc \textbf{72.3K} & \bc \textbf{460.77} \\
    \bottomrule
  \end{tabularx}

  \par\smallskip
  \begin{minipage}{\textwidth}
    \scriptsize
    \textsuperscript{*}Added terms report compression overhead.
  \end{minipage}
\end{table*}

\begin{figure}[htbp]
  \centering
  \includegraphics[width=0.92\columnwidth]{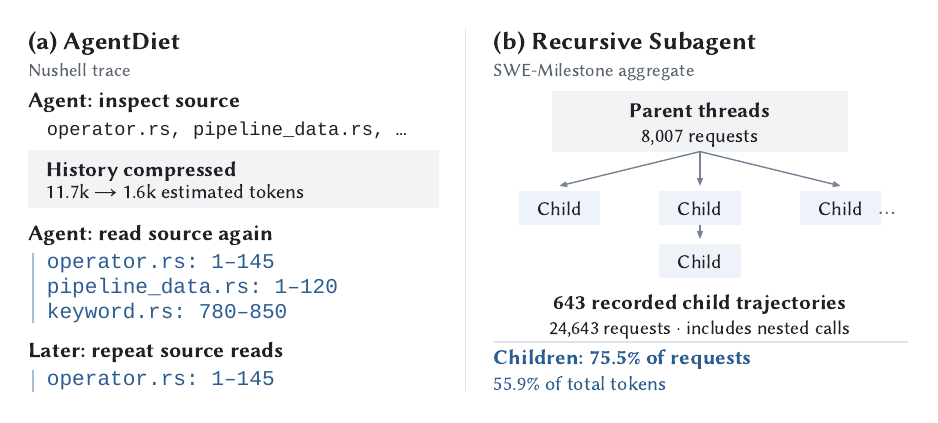}
  \caption{(a) Repeated source inspection with \agentdiet{}; (b) delegation with \recursivesubagent{}.}
  \label{fig:rq1-overhead}
\end{figure}

\modify{Table~\ref{tab:rq1-context-tools} presents the performance and cost comparison across context-management approaches on \swemilestone{}. In \agentdiet{}, reducing source-inspection history from 11.7k to 1.6k tokens while implementation is pending leads to repeated source rereading (Figure~\ref{fig:rq1-overhead}a), resolving fewer milestones than \basecodex{} (12 vs.\ 14) at higher cost (\$1,098.52 vs.\ \$681.22). Similarly, our \contextfolding{} baseline uses a single-level plan--branch--return workflow and resolves 13 milestones (vs.\ 14 for \basecodex{}), using 19.5\% more tokens at 29.4\% higher cost than \spinecodex{}. While \recursivesubagent{} achieves a Score comparable to \spinecodex{} (42.4\% vs.\ 42.6\%, resolving 17 vs.\ 19 milestones), establishing every subtask boundary through delegation causes child agents to account for 75.5\% of its 32,650 model requests (vs.\ 8,792 for \spinecodex{}; Figure~\ref{fig:rq1-overhead}b), driving cost to \$1,749.73 (vs.\ \$460.77). In contrast, \spinecodex{} executes nested subtasks locally while maintaining explicit work boundaries, resolving the most milestones (19) at the lowest cost (\$460.77).}

\subsubsection{RQ2: \modify{Generalizability} to \piagent{}}
\label{sec:eval-rq2}
\leavevmode\par

\ifdefined\resultmodelwidth\else
  \newlength{\resultmodelwidth}
  \newlength{\resultpanelwidth}
\fi
\begingroup
\scriptsize
\setlength{\tabcolsep}{2.5pt}
\settowidth{\resultmodelwidth}{\qwenthreeeighttwentysevenb}
\sbox0{%
  \begin{tabular}{@{}p{\resultmodelwidth}lrrrrrr@{}}
    \textbf{Model} & \textbf{Agent} & \textbf{Resolved} & \textbf{Almost}
      & \makecell[r]{\textbf{Avg.}\\\textbf{pass (\%)}}
      & \makecell[r]{\textbf{Total}\\\textbf{tokens}}
      & \makecell[r]{\textbf{Avg. input}\\\textbf{tokens/req.}}
      & \textbf{Cost (\$)} \\
    \qwenthreeeighttwentysevenb & \textbf{\spinepi{}} & 00/50 & 00/50 & 00.0 & 0.0M & 0.0K & 0 \\
  \end{tabular}%
}
\setlength{\resultpanelwidth}{\textwidth}
\global\resultmodelwidth=\resultmodelwidth
\global\resultpanelwidth=\resultpanelwidth
\endgroup

\begin{table}[t]
  \caption{ Performance of \spinepi{} and \basepi{} on \swemilestone{} and \programbench{}.}
  \label{tab:rq2-results}
  \centering
  \scriptsize
  \setlength{\tabcolsep}{2.5pt}
  \renewcommand{\arraystretch}{1.08}
  \newcommand{\bc}{\cellcolor{table_shallow_blue}}
  \newcommand{\tokhead}{\makecell[r]{\textbf{Total}\\\textbf{tokens}}}
  \newcommand{\avghead}{\makecell[r]{\textbf{Avg. input}\\\textbf{tokens/req.}}}
  \newcommand{\costhead}{\textbf{Cost (\$)}}

  \begin{tabularx}{\resultpanelwidth}{@{}p{\resultmodelwidth}Xrrrrr@{}}
    \toprule
    \multicolumn{7}{@{}l}{\textbf{\swemilestone{}}} \\
    \midrule
    \textbf{Model} & \textbf{Agent} & \textbf{Score (\%)} & \textbf{\# Resolved} & \tokhead & \avghead & \costhead \\
    \midrule
    \multirow{2}{\resultmodelwidth}{\geminithreeeightflash}
      & \bc \basepi{} & \bc 35.8 & \bc 13/98 & \bc 6,950.27M & \bc 511.06K & \bc 644.21 \\
      & \textbf{\spinepi{}} & \textbf{39.2} & \textbf{15/98} & \textbf{3,126.11M} & \textbf{164.31K} & \textbf{351.55} \\
    \midrule
    \multirow{2}{\resultmodelwidth}{\qwenthreeeighttwentysevenb}
      & \bc \basepi{} & \bc 30.9 & \bc 8/98 & \bc 1,996.97M & \bc 132.46K & \bc --- \\
      & \textbf{\spinepi{}} & \textbf{36.3} & \textbf{14/98} & \textbf{2,252.80M} & \textbf{148.73K} & \textbf{---} \\
    \bottomrule
  \end{tabularx}

  \vspace{6pt}
  \begin{tabularx}{\resultpanelwidth}{@{}p{\resultmodelwidth}Xrrrrrr@{}}
    \toprule
    \multicolumn{8}{@{}l}{\textbf{\programbench{}}} \\
    \midrule
    \textbf{Model} & \textbf{Agent} & \textbf{\# Resolved} & \textbf{\# Almost}
      & \makecell[r]{\textbf{Avg.}\\\textbf{pass (\%)}} & \tokhead & \avghead & \costhead \\
    \midrule
    \multirow{2}{\resultmodelwidth}{\geminithreeeightflash}
      & \bc \basepi{} & \bc 0/50 & \bc 5/50 & \bc 60.41 & \bc 1,663.4M & \bc 113.1K & \bc 217.08 \\
      & \textbf{\spinepi{}} & \textbf{1/50} & \textbf{9/50} & \textbf{65.65} & \textbf{1,419.2M} & \textbf{78.6K} & \textbf{200.96} \\
    \midrule
    \multirow{2}{\resultmodelwidth}{\qwenthreeeighttwentysevenb}
      & \bc \basepi{} & \bc 0/50 & \bc 2/50 & \bc 56.28 & \bc 239.84M & \bc 68.14K & \bc --- \\
      & \textbf{\spinepi{}} & \textbf{1/50} & \textbf{2/50} & \textbf{58.71} & \textbf{197.26M} & \textbf{51.63K} & \textbf{---} \\
    \bottomrule
  \end{tabularx}
\end{table}

In \spinepi{}, the agent declares work-scope boundaries through \piagent{}'s extension interface, and the runtime uses these declarations to prepare subsequent model inputs. \modify{Across both evaluated benchmarks, \spinepi{} improves task performance over \basepi{} (Table~\ref{tab:rq2-results}). On \swemilestone{}, \spinepi{} resolves more milestones than \basepi{} with both models, increasing the number from 13 to 15 with \geminithreeeightflash{} and from 8 to 14 with \qwenthreeeighttwentysevenb{}. The corresponding Score improves by 9.5\% (from 35.8\% to 39.2\%) with \geminithreeeightflash{} and by 17.5\% (from 30.9\% to 36.3\%) with \qwenthreeeighttwentysevenb{}. On \programbench{}, \spinepi{} fully resolves one task with each model, compared with none for \basepi{}, and increases the average test-pass rate by 8.7\% with \geminithreeeightflash{} (from 60.41\% to 65.65\%) and by 4.3\% with \qwenthreeeighttwentysevenb{} (from 56.28\% to 58.71\%).}

\modify{\spinepi{} also reduces average input tokens per request across most evaluated settings. With \geminithreeeightflash{}, average input drops by 67.8\% on \swemilestone{} (from 511.06K to 164.31K) and by 30.5\% on \programbench{}, cutting \swemilestone{} cost from \$644.21 to \$351.55 (45.4\%). This substantial reduction occurs because \basepi{} accumulates extensive flat history across Gemini's long window, whereas \spinepi{} replaces completed subtask details upon return. The exception to token reduction is \qwenthreeeighttwentysevenb{} on \swemilestone{}, where total tokens increase from 1,996.97M to 2,252.80M. Here, the baseline model frequently stalls or loses track of progress in accumulated history, whereas \spinepi{} preserves work boundaries, allowing the model to sustain execution and resolve six additional milestones.}

\findingbox{\textbf{Finding 2:} \modify{\spine{} generalizes effectively to \piagent{}, improving \swemilestone{} Score by 9.5--17.5\% and resolving more milestones across both models, while reducing average input tokens per request by up to 67.8\% and cutting commercial model cost by 45.4\%.}}%

\FloatBarrier
\subsubsection{RQ3: Behavioral Characteristics of \spine{}}
\label{sec:eval-rq3}
\leavevmode\par

\begin{figure}[htbp]
  \centering
  \includegraphics[width=0.95\columnwidth]{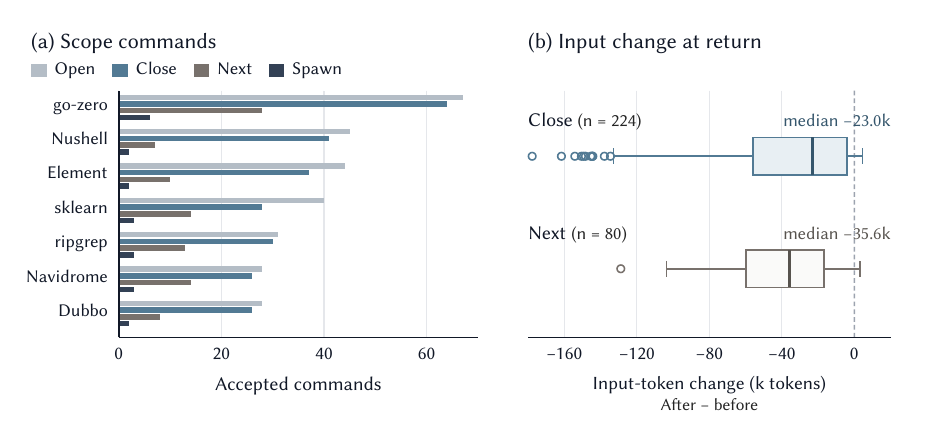}
  \caption{Runtime behavior of \spinecodex{}: (a) lifecycle calls; (b) input-token changes at returns.}
  \Description{Accepted Open, Close, Next, and Spawn calls across seven projects, and distributions of input-token changes at return boundaries.}
  \label{fig:rq3-runtime}
\end{figure}

\textbf{Runtime characteristics.} Agents use \spine{} to organize nested subtasks in seven \spinecodex{} project runs on \swemilestone{}, one per repository (Figure~\ref{fig:rq3-runtime}a). With the runtime root at depth zero, 72.6\% of requests with recorded scope depth occur at depths greater than one.

We also examine model input at return boundaries by comparing the inputs immediately before and after each accepted \closeop{} or \nextop{} call. The post-return input includes the returned result and other new messages, yet it is shorter after 223 of 224 \closeop{} calls and 77 of 80 \nextop{} calls, with median decreases across all calls of 23,045 and 35,627 tokens, respectively (Figure~\ref{fig:rq3-runtime}b). There, negative values indicate shorter input, and boxes show medians and interquartile ranges, with whiskers extending to observations within 1.5 interquartile ranges. Such measurements together show that agents use nested scopes beyond one level and return boundaries usually leave shorter inputs, consistent with the runtime replacing completed local interactions with returned results.

\begin{figure}[htbp]
  \centering
  \includegraphics[width=0.9\columnwidth]{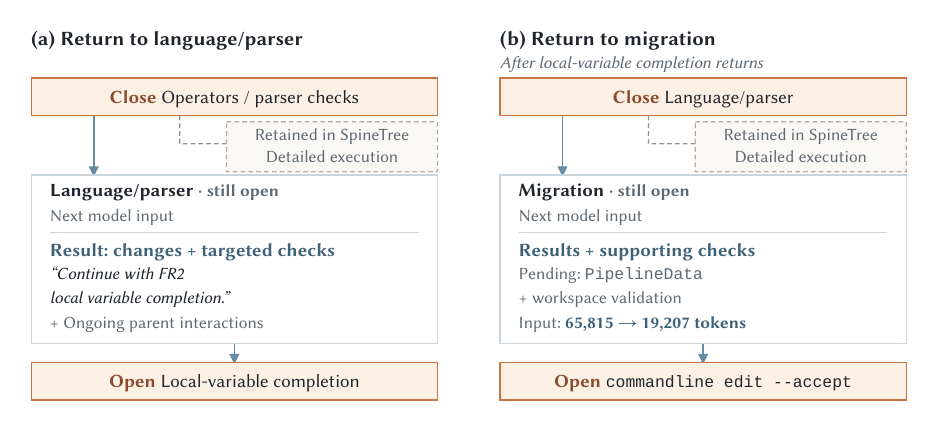}
  \caption{Nested returns during the \nushell{} migration with \spinecodex{}.}
  \Description{An inner scope returns to language/parser work, which later returns to the migration; each parent receives a result and continues pending work.}
  \label{fig:rq3-case}
\end{figure}

\textbf{Case study.} \modify{A recorded \spinecodex{} execution of the \nushell{} migration milestone (Section~\ref{sec:motivation}) illustrates how the agent manages subtask boundaries across nested scopes and continues from returned results (Figure~\ref{fig:rq3-case}). Within the language/parser update, the agent uses \openop{} to establish an inner scope for operator and parser checks. Upon passing targeted tests, the agent calls \closeop{} with a concise result; the runtime archives detailed execution in \spinetree{} and presents this result alongside ongoing parent context, allowing the same agent to directly continue with the pending local-variable completion (Figure~\ref{fig:rq3-case}a).}

\modify{When local-variable completion concludes, the agent calls \closeop{} on language/parser to return its aggregated result to the root migration. Across this outer return, model input drops from 65,815 to 19,207 tokens (70.8\%) as intermediate interactions recede into \spinetree{} while pending subtasks (\pipelinedata{} migration and workspace validation) remain clearly maintained in context (Figure~\ref{fig:rq3-case}b). By explicitly maintaining work states and context across boundaries, \spinecodex{} helps the agent sustain progress throughout the long-horizon task, ultimately resolving the migration milestone.}

\textbf{Ablation study.} \modify{To evaluate the core mechanisms in \modify{\spine{}}---subtask nesting and explicit work-boundary representation---we evaluate two ablations across three representative \swemilestone{} projects (Navidrome~\cite{swemilestone2026navidrome}, ripgrep~\cite{swemilestone2026ripgrep}, and Element~\cite{swemilestone2026element}) under our default model (\gptfivesixsol{} at high effort, three runs per configuration; Table~\ref{tab:rq3-navidrome}). \textbf{Depth 1} limits scopes to one level below the root, testing whether multi-level nesting is needed beyond single-level plan--branch--return workflows. \textbf{No boundary maintenance} removes explicit boundary markers and node states from model inputs while retaining runtime state compilation, isolating the effect of presenting boundary cues to the agent.}

\begin{table}[]
\centering
\caption{Ablation of nesting and boundary maintenance on three \swemilestone{} tasks.}
\label{tab:rq3-navidrome}
\small
\setlength{\tabcolsep}{3.5pt}
\scalebox{0.8}{
\begin{tabular}{lrrrrr}
\toprule
Configuration & Score (\%) & Resolved & Tokens & Avg. input & Cost \\
\midrule
\multicolumn{6}{l}{Navidrome, 9 milestones} \\
\spinecodex{} & 36.4 & 3/9 & 60.46M & 71.2K & \$42.48 \\
Depth 1 & 28.6 & 2.33/9 & 71.50M & 99.5K & \$52.60 \\
No boundary maintenance & 29.0 & 2.33/9 & 67.40M & 124.2K & \$60.50 \\
\midrule
\multicolumn{6}{l}{ripgrep, 11 milestones} \\
\spinecodex{} & 72.4 & 5.667/11 & 48.22M & 59.8K & \$34.67 \\
Depth 1 & 66.8 & 4/11 & 54.43M & 82.8K & \$40.78 \\
No boundary maintenance & 67.2 & 3/11 & 52.77M & 118.1K & \$45.88 \\
\midrule
\multicolumn{6}{l}{Element, 18 milestones} \\
\spinecodex{} & 33.7 & 2.667/18 & 112.12M & 78.1K & \$78.87 \\
Depth 1 & 27.4 & 1.667/18 & 123.60M & 98.7K & \$91.79 \\
No boundary maintenance & 27.5 & 1.333/18 & 114.10M & 128.6K & \$102.24 \\
\bottomrule
\end{tabular}
}
\end{table}

\FloatBarrier

\begin{samepage}

\modify{Across all three projects, \spinecodex{} improves Score by 7.7--27.3\% over both ablations, resolving more milestones at lower cost with shorter inputs (Table~\ref{tab:rq3-navidrome}). Restricting nesting (Depth 1) raises average input per request from 59.8K--78.1K to 82.8K--99.5K, showing that shallow scopes fail to isolate intermediate interactions within nested subtasks. More prominently, removing boundary cues causes average input to surge to 118.1K--128.6K (+64--97\%) and cost to rise by 30--42\%. This confirms the first gap in Section~\ref{sec:introduction}: without explicit boundaries and completion states in context, the agent may lose track of active scopes, accumulating redundant interactions and delaying returns.}%

\end{samepage}

\section{Threats to Validity}
\label{sec:threats}

\noindent\textbf{Threats to internal validity.}
\modify{%
Differences in experimental settings and implementation choices may affect comparisons between methods.
To mitigate these effects, we evaluate all approaches under the same model, harness, task set, and execution budget within each comparison, and conduct all runs in isolated containers.
Furthermore, to ensure that observed improvements stem from the core mechanisms of \spine{} rather than incidental factors, we conduct an ablation study (Section~\ref{sec:eval-rq3}) that isolates scope nesting and explicit work-boundary representation, confirming their individual contributions to both task effectiveness and execution cost.}

\noindent\textbf{Threats to external validity.}
The generalizability of our findings may depend on the tasks, models, and agent harnesses selected for evaluation. To address this, we evaluate \spine{} on three benchmarks covering repository evolution, whole-program reconstruction, and open-ended engineering. We also integrate \spine{} into \codex{} and \piagent{} and evaluate these integrations using multiple models, including an open-weight model, to examine its applicability across different tasks and systems.

\noindent\textbf{Threats to construct validity.}
A single metric may not capture both task completion and execution cost. To address this, we adopt each benchmark's native evaluation metrics to measure task completion and partial progress, and complement them with token usage and monetary cost to assess execution overhead.

\section{Conclusion}
\label{sec:conclusion}

We presented \spine{}, an interface that enables agents to explicitly declare subtask boundaries, allowing model context to align with subtask nesting and completion. It separates subtask boundaries from delegation and uses \spinetree{} to construct context that retains unfinished subtasks' interactions and replaces completed subtasks' details with their results. We integrated \spine{} into \codex{} and evaluated it on \swemilestone{}, \programbench{}, and \frontierswe{}. With \gptfivesixsol{}, \spinecodex{} improves \swemilestone{} Score by 11.8\% while reducing total cost by 32.4\%, compared with native \codex{}. A plugin integration into \piagent{} also improves \swemilestone{} Score and the average \programbench{} test-pass rate compared with native \piagent{} under both evaluated models, including an open-weight model. By decoupling subtask boundaries from delegation and compiling context around hierarchical work structures just in time, \spine{} demonstrates that coding agents can sustain complex, long-horizon engineering tasks effectively and economically.

\section*{Data Availability}
\spinecodex{} and \spinepi{} are open source on GitHub~\cite{spinecodexrepo,spinepirepo} and published on npm, ready to use out of the box:
\begin{itemize}
  \item \spinecodex{}: \texttt{npm install -g @spinejit/spine-codex}, then run \texttt{spine-codex};
  \item \spinepi{}: \texttt{pi install npm:@spinejit/pi-spinejit} in an existing \piagent{} installation.
\end{itemize}

\newpage

\bibliographystyle{ACM-Reference-Format}
\bibliography{references}


\begin{thebibliography}{62}


\ifx \showCODEN    \undefined \def \showCODEN     #1{\unskip}     \fi
\ifx \showISBNx    \undefined \def \showISBNx     #1{\unskip}     \fi
\ifx \showISBNxiii \undefined \def \showISBNxiii  #1{\unskip}     \fi
\ifx \showISSN     \undefined \def \showISSN      #1{\unskip}     \fi
\ifx \showLCCN     \undefined \def \showLCCN      #1{\unskip}     \fi
\ifx \shownote     \undefined \def \shownote      #1{#1}          \fi
\ifx \showarticletitle \undefined \def \showarticletitle #1{#1}   \fi
\ifx \showURL      \undefined \def \showURL       {\relax}        \fi
\providecommand\bibfield[2]{#2}
\providecommand\bibinfo[2]{#2}
\providecommand\natexlab[1]{#1}
\providecommand\showeprint[2][]{arXiv:#2}
\makeatletter
\@ifundefined{NAT@parse@date}{}{\let\NAT@parse@date@orig\NAT@parse@date}
\@ifundefined{NAT@parse@date}{}{\def\NAT@parse@date#1#2#3#4#5#6@@{\NAT@parse@date@orig#1#2#3#4#5#6@@\def\NAT@tempyear{0000}\def\NAT@tempexlab{{?}}\ifx\NAT@year\NAT@tempyear\ifx\NAT@exlab\NAT@tempexlab\def\NAT@date{[n.\,d.]}\else\edef\NAT@date{[n.\,d.]\NAT@exlab}\fi\fi}}
\makeatother

\bibitem[Aho and Johnson(1974)]%
        {aho1974lrparsing}
\bibfield{author}{\bibinfo{person}{A.~V. Aho} {and} \bibinfo{person}{S.~C.
  Johnson}.} \bibinfo{year}{1974}\natexlab{}.
\newblock \showarticletitle{{LR} Parsing}.
\newblock \bibinfo{journal}{\emph{Comput. Surveys}} \bibinfo{volume}{6},
  \bibinfo{number}{2} (\bibinfo{date}{June} \bibinfo{year}{1974}),
  \bibinfo{pages}{99--124}.
\newblock
\href{https://doi.org/10.1145/356628.356629}{doi:\nolinkurl{10.1145/356628.356629}}


\bibitem[{Anomalyco}(2026)]%
        {opencode2026repo}
\bibfield{author}{\bibinfo{person}{{Anomalyco}}.}
  \bibinfo{year}{2026}\natexlab{}.
\newblock \bibinfo{title}{{OpenCode}: The Open Source {AI} Coding Agent}.
\newblock \bibinfo{howpublished}{GitHub repository}.
\newblock
\urldef\tempurl%
\url{https://github.com/anomalyco/opencode}
\showURL{%
\tempurl}


\bibitem[{Anthropic}(0000)]%
        {anthropicRakuten}
\bibfield{author}{\bibinfo{person}{{Anthropic}}.}
  \bibinfo{year}{[n.\,d.]}\natexlab{}.
\newblock \bibinfo{title}{{Rakuten} Accelerates Development with {Claude
  Code}}.
\newblock \bibinfo{howpublished}{Customer case study}.
\newblock
\urldef\tempurl%
\url{https://www.anthropic.com/customers/rakuten}
\showURL{%
\tempurl}
\newblock
\shownote{Accessed October 2, 2026}.


\bibitem[{Anthropic}(2026a)]%
        {Anthropic-ClaudeCode-DynamicWorkflows}
\bibfield{author}{\bibinfo{person}{{Anthropic}}.}
  \bibinfo{year}{2026}\natexlab{a}.
\newblock \bibinfo{title}{Orchestrate Subagents at Scale with Dynamic
  Workflows}.
\newblock \bibinfo{howpublished}{Claude Code documentation}.
\newblock
\urldef\tempurl%
\url{https://code.claude.com/docs/en/workflows}
\showURL{%
\tempurl}


\bibitem[{Anthropic}(2026b)]%
        {anthropic2026managedagents}
\bibfield{author}{\bibinfo{person}{{Anthropic}}.}
  \bibinfo{year}{2026}\natexlab{b}.
\newblock \bibinfo{title}{Scaling Managed Agents: Decoupling the Brain from the
  Hands}.
\newblock \bibinfo{howpublished}{Anthropic Engineering}.
\newblock
\urldef\tempurl%
\url{https://www.anthropic.com/engineering/managed-agents}
\showURL{%
\tempurl}


\bibitem[Chu et~al\mbox{.}(2026)]%
        {proximal2026frontierswe}
\bibfield{author}{\bibinfo{person}{Evan Chu}, \bibinfo{person}{Rajan Agarwal},
  \bibinfo{person}{Abishek Thangamuthu}, \bibinfo{person}{Brendan Graham},
  \bibinfo{person}{Justus Mattern}, \bibinfo{person}{Freeman Jiang},
  \bibinfo{person}{Paul Cento}, \bibinfo{person}{Swarnim Jain},
  \bibinfo{person}{Mersad Abbasi}, \bibinfo{person}{Mohammad~Hossein Rezaei},
  \bibinfo{person}{George Wang}, \bibinfo{person}{Alex Zhang},
  \bibinfo{person}{Simon Guo}, \bibinfo{person}{Karina Nguyen},
  \bibinfo{person}{Danna Liu}, \bibinfo{person}{Arash Bidgoli},
  \bibinfo{person}{Aditya Dalmia}, \bibinfo{person}{Apoorv Dankar},
  \bibinfo{person}{Ashrut Vaddela}, \bibinfo{person}{Calvin Chen},
  \bibinfo{person}{Keshav Kumar}, \bibinfo{person}{Kushagra Vaish},
  \bibinfo{person}{Navid Pour}, \bibinfo{person}{Rishyanth Kondra},
  \bibinfo{person}{Sagar Badiyani}, \bibinfo{person}{Sidharth Giri},
  \bibinfo{person}{Snagnik Das}, \bibinfo{person}{Soham Gaikwad},
  \bibinfo{person}{Syed Shah}, \bibinfo{person}{Vagish Dilawari}, {and}
  \bibinfo{person}{Vishal Agarwal}.} \bibinfo{year}{2026}\natexlab{}.
\newblock \bibinfo{title}{{FrontierSWE}}.
\newblock \bibinfo{howpublished}{Proximal Blog}.
\newblock
\urldef\tempurl%
\url{https://www.frontierswe.com/blog/v1}
\showURL{%
\tempurl}


\bibitem[{Cursor Team}(2026)]%
        {cursor2026longrunning}
\bibfield{author}{\bibinfo{person}{{Cursor Team}}.}
  \bibinfo{year}{2026}\natexlab{}.
\newblock \bibinfo{title}{Expanding our Long-Running Agents Research Preview}.
\newblock \bibinfo{howpublished}{Cursor Engineering Blog}.
\newblock
\urldef\tempurl%
\url{https://cursor.com/blog/long-running-agents}
\showURL{%
\tempurl}


\bibitem[{DeepCommit-ai}(2026a)]%
        {swemilestone2026element}
\bibfield{author}{\bibinfo{person}{{DeepCommit-ai}}.}
  \bibinfo{year}{2026}\natexlab{a}.
\newblock \bibinfo{title}{{SWE-Milestone}: {Element Web} Task Data}.
\newblock \bibinfo{howpublished}{Hugging Face dataset}.
\newblock
\urldef\tempurl%
\url{https://huggingface.co/datasets/DeepCommit-ai/SWE-Milestone-data/tree/3500478e740d42583277c2658aa7729eb36e10f4/element-hq_element-web_v1.11.95_v1.11.97}
\showURL{%
\tempurl}
\newblock
\shownote{Release range: v1.11.95--v1.11.97}.


\bibitem[{DeepCommit-ai}(2026b)]%
        {swemilestone2026protocol}
\bibfield{author}{\bibinfo{person}{{DeepCommit-ai}}.}
  \bibinfo{year}{2026}\natexlab{b}.
\newblock \bibinfo{title}{{SWE-Milestone}: Evaluation Protocol}.
\newblock \bibinfo{howpublished}{Official benchmark repository README,
  Troubleshooting}.
\newblock
\urldef\tempurl%
\url{https://github.com/DeepCommit-ai/SWE-Milestone/blob/17a8f1593e172e26b36cea15e2b30fb9536c93f5/README.md}
\showURL{%
\tempurl}


\bibitem[{DeepCommit-ai}(2026c)]%
        {swemilestone2026navidrome}
\bibfield{author}{\bibinfo{person}{{DeepCommit-ai}}.}
  \bibinfo{year}{2026}\natexlab{c}.
\newblock \bibinfo{title}{{SWE-Milestone}: {Navidrome} Task Data}.
\newblock \bibinfo{howpublished}{Hugging Face dataset}.
\newblock
\urldef\tempurl%
\url{https://huggingface.co/datasets/DeepCommit-ai/SWE-Milestone-data/tree/3500478e740d42583277c2658aa7729eb36e10f4/navidrome_navidrome_v0.57.0_v0.58.0}
\showURL{%
\tempurl}
\newblock
\shownote{Release range: v0.57.0--v0.58.0}.


\bibitem[{DeepCommit-ai}(2026d)]%
        {swemilestone2026ripgrep}
\bibfield{author}{\bibinfo{person}{{DeepCommit-ai}}.}
  \bibinfo{year}{2026}\natexlab{d}.
\newblock \bibinfo{title}{{SWE-Milestone}: {ripgrep} Task Data}.
\newblock \bibinfo{howpublished}{Hugging Face dataset}.
\newblock
\urldef\tempurl%
\url{https://huggingface.co/datasets/DeepCommit-ai/SWE-Milestone-data/tree/3500478e740d42583277c2658aa7729eb36e10f4/BurntSushi_ripgrep_14.1.1_15.0.0}
\showURL{%
\tempurl}
\newblock
\shownote{Release range: 14.1.1--15.0.0}.


\bibitem[{DeepSeek AI}(2026)]%
        {deepseek2026harness}
\bibfield{author}{\bibinfo{person}{{DeepSeek AI}}.}
  \bibinfo{year}{2026}\natexlab{}.
\newblock \bibinfo{title}{{DeepSeek Harness}: Everything is a Plugin}.
\newblock \bibinfo{howpublished}{GitHub repository}.
\newblock
\urldef\tempurl%
\url{https://github.com/deepseek-ai/deepseek-harness}
\showURL{%
\tempurl}


\bibitem[Deng et~al\mbox{.}(2026)]%
        {deng2026evoclaw}
\bibfield{author}{\bibinfo{person}{Gangda Deng}, \bibinfo{person}{Zhaoling
  Chen}, \bibinfo{person}{Zhongming Yu}, \bibinfo{person}{Haoyang Fan},
  \bibinfo{person}{Yuhong Liu}, \bibinfo{person}{Yuxin Yang},
  \bibinfo{person}{Dhruv Parikh}, \bibinfo{person}{Rajgopal Kannan},
  \bibinfo{person}{Le Cong}, \bibinfo{person}{Mengdi Wang},
  \bibinfo{person}{Qian Zhang}, \bibinfo{person}{Viktor Prasanna},
  \bibinfo{person}{Xiangru Tang}, {and} \bibinfo{person}{Xingyao Wang}.}
  \bibinfo{year}{2026}\natexlab{}.
\newblock \bibinfo{title}{{SWE-Milestone}: Evaluating {AI} Agents on Continuous
  Software Evolution}.
\newblock
\showeprint[arxiv]{2603.13428}~[cs.SE]
\urldef\tempurl%
\url{https://arxiv.org/abs/2603.13428}
\showURL{%
\tempurl}


\bibitem[{Earendil}(2026)]%
        {pi2026npm}
\bibfield{author}{\bibinfo{person}{{Earendil}}.}
  \bibinfo{year}{2026}\natexlab{}.
\newblock \bibinfo{title}{{@earendil-works/pi-coding-agent} npm package}.
\newblock \bibinfo{howpublished}{npm registry; weekly-download snapshot for
  2026-09-24--2026-09-30}.
\newblock
\urldef\tempurl%
\url{https://www.npmjs.com/package/@earendil-works/pi-coding-agent}
\showURL{%
\tempurl}


\bibitem[{earendil-works}(2026a)]%
        {earendil2026pirepo}
\bibfield{author}{\bibinfo{person}{{earendil-works}}.}
  \bibinfo{year}{2026}\natexlab{a}.
\newblock \bibinfo{title}{{Pi}: {AI} Agent Toolkit, Unified {LLM} {API}, Agent
  Loop, {TUI}, and Coding Agent {CLI}}.
\newblock \bibinfo{howpublished}{GitHub repository}.
\newblock
\urldef\tempurl%
\url{https://github.com/earendil-works/pi}
\showURL{%
\tempurl}


\bibitem[{earendil-works}(2026b)]%
        {earendil2026piextensions}
\bibfield{author}{\bibinfo{person}{{earendil-works}}.}
  \bibinfo{year}{2026}\natexlab{b}.
\newblock \bibinfo{title}{{Pi} Extensions}.
\newblock \bibinfo{howpublished}{Pi documentation}.
\newblock
\urldef\tempurl%
\url{https://github.com/earendil-works/pi/blob/main/packages/coding-agent/docs/extensions.md}
\showURL{%
\tempurl}


\bibitem[Gao et~al\mbox{.}(2026)]%
        {gao2026swemem}
\bibfield{author}{\bibinfo{person}{Shuzheng Gao}, \bibinfo{person}{Wenhao
  Zeng}, \bibinfo{person}{Zhaojian Yu}, \bibinfo{person}{Jianqiao Wangni},
  \bibinfo{person}{Chaozheng Wang}, \bibinfo{person}{Kai Cai},
  \bibinfo{person}{Shilin He}, {and} \bibinfo{person}{Michael~R. Lyu}.}
  \bibinfo{year}{2026}\natexlab{}.
\newblock \bibinfo{title}{{SWE-MeM}: Learning Adaptive Memory Management for
  Long-Horizon Coding Agents}.
\newblock
\showeprint[arxiv]{2606.28434}~[cs.SE]
\urldef\tempurl%
\url{https://arxiv.org/abs/2606.28434}
\showURL{%
\tempurl}


\bibitem[{GhabiX}(2026a)]%
        {spinepirepo}
\bibfield{author}{\bibinfo{person}{{GhabiX}}.}
  \bibinfo{year}{2026}\natexlab{a}.
\newblock \bibinfo{title}{{spine-plugin}}.
\newblock \bibinfo{howpublished}{GitHub repository}.
\newblock
\urldef\tempurl%
\url{https://github.com/GhabiX/spine-plugin}
\showURL{%
\tempurl}


\bibitem[{GhabiX}(2026b)]%
        {spinecodexrepo}
\bibfield{author}{\bibinfo{person}{{GhabiX}}.}
  \bibinfo{year}{2026}\natexlab{b}.
\newblock \bibinfo{title}{{SpineCodex}}.
\newblock \bibinfo{howpublished}{GitHub repository}.
\newblock
\urldef\tempurl%
\url{https://github.com/GhabiX/SpineCodex}
\showURL{%
\tempurl}


\bibitem[{Google}(2026)]%
        {google2026gemini38flash}
\bibfield{author}{\bibinfo{person}{{Google}}.} \bibinfo{year}{2026}\natexlab{}.
\newblock \bibinfo{title}{{Gemini} 3.8 {Flash}}.
\newblock \bibinfo{howpublished}{Gemini API documentation}.
\newblock
\urldef\tempurl%
\url{https://ai.google.dev/gemini-api/docs/models/gemini-3.8-flash}
\showURL{%
\tempurl}


\bibitem[Hu et~al\mbox{.}(2025)]%
        {hu2025hiagent}
\bibfield{author}{\bibinfo{person}{Mengkang Hu}, \bibinfo{person}{Tianxing
  Chen}, \bibinfo{person}{Qiguang Chen}, \bibinfo{person}{Yao Mu},
  \bibinfo{person}{Wenqi Shao}, {and} \bibinfo{person}{Ping Luo}.}
  \bibinfo{year}{2025}\natexlab{}.
\newblock \showarticletitle{{HiAgent}: Hierarchical Working Memory Management
  for Solving Long-Horizon Agent Tasks with Large Language Model}. In
  \bibinfo{booktitle}{\emph{Proceedings of the 63rd Annual Meeting of the
  Association for Computational Linguistics}}. \bibinfo{pages}{32779--32798}.
\newblock
\urldef\tempurl%
\url{https://aclanthology.org/2025.acl-long.1575/}
\showURL{%
\tempurl}


\bibitem[Johnston et~al\mbox{.}(2026)]%
        {codex2026agenticwork}
\bibfield{author}{\bibinfo{person}{Drew Johnston}, \bibinfo{person}{David
  Holtz}, \bibinfo{person}{Alex~Martin Richmond}, \bibinfo{person}{Christopher
  Ong}, \bibinfo{person}{Prasanna Tambe}, {and} \bibinfo{person}{Aaron
  Chatterji}.} \bibinfo{year}{2026}\natexlab{}.
\newblock \bibinfo{title}{The Shift to Agentic {AI}: Evidence from {Codex}}.
\newblock
\showeprint[arxiv]{2606.26959}~[econ.GN]
\urldef\tempurl%
\url{https://arxiv.org/abs/2606.26959}
\showURL{%
\tempurl}


\bibitem[Kang et~al\mbox{.}(2026)]%
        {kang2026acon}
\bibfield{author}{\bibinfo{person}{Minki Kang}, \bibinfo{person}{Wei-Ning
  Chen}, \bibinfo{person}{Dongge Han}, \bibinfo{person}{Huseyin~A. Inan},
  \bibinfo{person}{Lukas Wutschitz}, \bibinfo{person}{Yanzhi Chen},
  \bibinfo{person}{Robert Sim}, {and} \bibinfo{person}{Saravan Rajmohan}.}
  \bibinfo{year}{2026}\natexlab{}.
\newblock \showarticletitle{{ACON}: Optimizing Context Compression for
  Long-horizon {LLM} Agents}. In \bibinfo{booktitle}{\emph{Proceedings of the
  43rd International Conference on Machine Learning}}
  \emph{(\bibinfo{series}{Proceedings of Machine Learning Research},
  Vol.~\bibinfo{volume}{306})}. \bibinfo{pages}{55321--55375}.
\newblock
\urldef\tempurl%
\url{https://proceedings.mlr.press/v306/kang26b.html}
\showURL{%
\tempurl}


\bibitem[Kwon et~al\mbox{.}(2023)]%
        {kwon2023efficient}
\bibfield{author}{\bibinfo{person}{Woosuk Kwon}, \bibinfo{person}{Zhuohan Li},
  \bibinfo{person}{Siyuan Zhuang}, \bibinfo{person}{Ying Sheng},
  \bibinfo{person}{Lianmin Zheng}, \bibinfo{person}{Cody~Hao Yu},
  \bibinfo{person}{Joseph~E. Gonzalez}, \bibinfo{person}{Hao Zhang}, {and}
  \bibinfo{person}{Ion Stoica}.} \bibinfo{year}{2023}\natexlab{}.
\newblock \showarticletitle{Efficient Memory Management for Large Language
  Model Serving with {PagedAttention}}. In
  \bibinfo{booktitle}{\emph{Proceedings of the 29th Symposium on Operating
  Systems Principles}}. \bibinfo{pages}{611--626}.
\newblock
\href{https://doi.org/10.1145/3600006.3613165}{doi:\nolinkurl{10.1145/3600006.3613165}}


\bibitem[Li et~al\mbox{.}(2026)]%
        {li2026sculptor}
\bibfield{author}{\bibinfo{person}{Mo Li}, \bibinfo{person}{L.~H. Xu},
  \bibinfo{person}{Qitai Tan}, \bibinfo{person}{Long Ma},
  \bibinfo{person}{Flood Sung}, \bibinfo{person}{Ting Cao}, {and}
  \bibinfo{person}{Yunxin Liu}.} \bibinfo{year}{2026}\natexlab{}.
\newblock \showarticletitle{{Sculptor}: Empowering {LLM}s with Cognitive Agency
  via Active Context Management}. In \bibinfo{booktitle}{\emph{International
  Conference on Learning Representations}}.
\newblock
\urldef\tempurl%
\url{https://proceedings.iclr.cc/paper_files/paper/2026/file/f9155a06ba1b20e40c38bf8541db456f-Paper-Conference.pdf}
\showURL{%
\tempurl}


\bibitem[Liu et~al\mbox{.}(2026)]%
        {liu2026contexttool}
\bibfield{author}{\bibinfo{person}{Shukai Liu}, \bibinfo{person}{Bo Jiang},
  \bibinfo{person}{Jian Yang}, \bibinfo{person}{Yizhi Li},
  \bibinfo{person}{Jinyang Guo}, \bibinfo{person}{Xianglong Liu}, {and}
  \bibinfo{person}{Bryan Dai}.} \bibinfo{year}{2026}\natexlab{}.
\newblock \showarticletitle{Context as a Tool: Context Management for
  Long-Horizon {SWE}-Agents}. In \bibinfo{booktitle}{\emph{Findings of the
  Association for Computational Linguistics: ACL 2026}}.
  \bibinfo{publisher}{Association for Computational Linguistics},
  \bibinfo{address}{San Diego, California, United States},
  \bibinfo{pages}{20604--20617}.
\newblock
\href{https://doi.org/10.18653/v1/2026.findings-acl.1032}{doi:\nolinkurl{10.18653/v1/2026.findings-acl.1032}}


\bibitem[Lumer et~al\mbox{.}(2026a)]%
        {dontbreakcache2026}
\bibfield{author}{\bibinfo{person}{Elias Lumer}, \bibinfo{person}{Faheem
  Nizar}, \bibinfo{person}{Akshaya Jangiti}, \bibinfo{person}{Kevin Frank},
  \bibinfo{person}{Anmol Gulati}, \bibinfo{person}{Mandar Phadate}, {and}
  \bibinfo{person}{Vamse~Kumar Subbiah}.} \bibinfo{year}{2026}\natexlab{a}.
\newblock \bibinfo{title}{Don't Break the Cache: An Evaluation of Prompt
  Caching for Long-Horizon Agentic Tasks}.
\newblock
\showeprint[arxiv]{2601.06007}
\urldef\tempurl%
\url{https://arxiv.org/abs/2601.06007}
\showURL{%
\tempurl}


\bibitem[Lumer et~al\mbox{.}(2026b)]%
        {rah2026}
\bibfield{author}{\bibinfo{person}{Elias Lumer}, \bibinfo{person}{Sahil Sen},
  \bibinfo{person}{Kevin Paul}, {and} \bibinfo{person}{Vamse~Kumar Subbiah}.}
  \bibinfo{year}{2026}\natexlab{b}.
\newblock \bibinfo{title}{Recursive Agent Harnesses}.
\newblock
\showeprint[arxiv]{2606.13643}
\urldef\tempurl%
\url{https://arxiv.org/abs/2606.13643}
\showURL{%
\tempurl}


\bibitem[{OpenAI}(2025)]%
        {openai2025codexmax}
\bibfield{author}{\bibinfo{person}{{OpenAI}}.} \bibinfo{year}{2025}\natexlab{}.
\newblock \bibinfo{title}{Building More with {GPT}-5.1-{Codex}-{Max}}.
\newblock \bibinfo{howpublished}{OpenAI}.
\newblock
\urldef\tempurl%
\url{https://openai.com/index/gpt-5-1-codex-max/}
\showURL{%
\tempurl}


\bibitem[{OpenAI}(2026a)]%
        {openai2026codexknowledgework}
\bibfield{author}{\bibinfo{person}{{OpenAI}}.}
  \bibinfo{year}{2026}\natexlab{a}.
\newblock \bibinfo{title}{Codex Is Becoming a Productivity Tool for Everyone}.
\newblock \bibinfo{howpublished}{OpenAI}.
\newblock
\urldef\tempurl%
\url{https://openai.com/index/codex-for-knowledge-work/}
\showURL{%
\tempurl}


\bibitem[{OpenAI}(2026b)]%
        {openai2026codexrepo}
\bibfield{author}{\bibinfo{person}{{OpenAI}}.}
  \bibinfo{year}{2026}\natexlab{b}.
\newblock \bibinfo{title}{{Codex}: Lightweight Coding Agent that Runs in Your
  Terminal}.
\newblock \bibinfo{howpublished}{GitHub repository}.
\newblock
\urldef\tempurl%
\url{https://github.com/openai/codex}
\showURL{%
\tempurl}


\bibitem[{OpenAI}(2026c)]%
        {openai2026codexmultiagentv2}
\bibfield{author}{\bibinfo{person}{{OpenAI}}.}
  \bibinfo{year}{2026}\natexlab{c}.
\newblock \bibinfo{title}{{Codex} Multi-Agent V2 Spawn Handler}.
\newblock \bibinfo{howpublished}{GitHub source code}.
\newblock
\urldef\tempurl%
\url{https://github.com/openai/codex/blob/main/codex-rs/core/src/tools/handlers/multi_agents_v2/spawn.rs}
\showURL{%
\tempurl}


\bibitem[{OpenAI}(2026d)]%
        {openai2026gpt56sol}
\bibfield{author}{\bibinfo{person}{{OpenAI}}.}
  \bibinfo{year}{2026}\natexlab{d}.
\newblock \bibinfo{title}{{GPT}-5.6 {Sol} Model}.
\newblock \bibinfo{howpublished}{OpenAI API documentation}.
\newblock
\urldef\tempurl%
\url{https://developers.openai.com/api/docs/models/gpt-5.6-sol}
\showURL{%
\tempurl}


\bibitem[{ProgramBench}(2026a)]%
        {programbench2026ffmpeg}
\bibfield{author}{\bibinfo{person}{{ProgramBench}}.}
  \bibinfo{year}{2026}\natexlab{a}.
\newblock \bibinfo{title}{{ProgramBench}: {FFmpeg}/{FFmpeg} Task}.
\newblock \bibinfo{howpublished}{Official benchmark task page}.
\newblock
\urldef\tempurl%
\url{https://programbench.com/task/ffmpeg__ffmpeg.360a402/}
\showURL{%
\tempurl}


\bibitem[{ProgramBench}(2026b)]%
        {programbench2026sqlite}
\bibfield{author}{\bibinfo{person}{{ProgramBench}}.}
  \bibinfo{year}{2026}\natexlab{b}.
\newblock \bibinfo{title}{{ProgramBench}: {SQLite}/{SQLite} Task}.
\newblock \bibinfo{howpublished}{Official benchmark task page}.
\newblock
\urldef\tempurl%
\url{https://programbench.com/task/sqlite__sqlite.839433d/}
\showURL{%
\tempurl}


\bibitem[{Proximal}(2026a)]%
        {frontierswe2026libswscale}
\bibfield{author}{\bibinfo{person}{{Proximal}}.}
  \bibinfo{year}{2026}\natexlab{a}.
\newblock \bibinfo{title}{{FrontierSWE} V1: {FFmpeg} libswscale
  Re-implementation}.
\newblock \bibinfo{howpublished}{FrontierSWE V1 task page}.
\newblock
\urldef\tempurl%
\url{https://www.frontierswe.com/v1/ffmpeg-swscale-rewrite}
\showURL{%
\tempurl}


\bibitem[{Proximal}(2026b)]%
        {frontierswe2026frogsgame}
\bibfield{author}{\bibinfo{person}{{Proximal}}.}
  \bibinfo{year}{2026}\natexlab{b}.
\newblock \bibinfo{title}{{FrontierSWE} V1: FrogsGame Post-Training}.
\newblock \bibinfo{howpublished}{Official benchmark task page}.
\newblock
\urldef\tempurl%
\url{https://www.frontierswe.com/v1/frogsgame-rl}
\showURL{%
\tempurl}


\bibitem[{Proximal}(2026c)]%
        {frontierswe2026gittozig}
\bibfield{author}{\bibinfo{person}{{Proximal}}.}
  \bibinfo{year}{2026}\natexlab{c}.
\newblock \bibinfo{title}{{FrontierSWE} V1: {Git} to {Zig}}.
\newblock \bibinfo{howpublished}{FrontierSWE V1 task page}.
\newblock
\urldef\tempurl%
\url{https://www.frontierswe.com/v1/git-to-zig}
\showURL{%
\tempurl}


\bibitem[{Proximal}(2026d)]%
        {frontierswe2026optimizer}
\bibfield{author}{\bibinfo{person}{{Proximal}}.}
  \bibinfo{year}{2026}\natexlab{d}.
\newblock \bibinfo{title}{{FrontierSWE} V1: Optimizer Design}.
\newblock \bibinfo{howpublished}{Official benchmark task page}.
\newblock
\urldef\tempurl%
\url{https://www.frontierswe.com/v1/optimizer-design}
\showURL{%
\tempurl}


\bibitem[{Proximal}(2026e)]%
        {frontierswe2026postgresqlsqlite}
\bibfield{author}{\bibinfo{person}{{Proximal}}.}
  \bibinfo{year}{2026}\natexlab{e}.
\newblock \bibinfo{title}{{FrontierSWE} V1: {PostgreSQL} 18 on {SQLite}}.
\newblock \bibinfo{howpublished}{FrontierSWE V1 task page}.
\newblock
\urldef\tempurl%
\url{https://www.frontierswe.com/v1/postgres-sqlite-wire-adapter}
\showURL{%
\tempurl}


\bibitem[{Qwen Team}(2026)]%
        {qwen2026qwen38}
\bibfield{author}{\bibinfo{person}{{Qwen Team}}.}
  \bibinfo{year}{2026}\natexlab{}.
\newblock \bibinfo{title}{{Qwen3.8}-27B}.
\newblock \bibinfo{howpublished}{Hugging Face model card}.
\newblock
\urldef\tempurl%
\url{https://huggingface.co/Qwen/Qwen3.8-27B}
\showURL{%
\tempurl}


\bibitem[Schroeder et~al\mbox{.}(2025)]%
        {schroeder2025thread}
\bibfield{author}{\bibinfo{person}{Philip Schroeder},
  \bibinfo{person}{Nathaniel~W. Morgan}, \bibinfo{person}{Hongyin Luo}, {and}
  \bibinfo{person}{James Glass}.} \bibinfo{year}{2025}\natexlab{}.
\newblock \showarticletitle{{THREAD}: Thinking Deeper with Recursive Spawning}.
  In \bibinfo{booktitle}{\emph{Proceedings of the 2025 Conference of the
  Nations of the Americas Chapter of the Association for Computational
  Linguistics: Human Language Technologies}}. \bibinfo{pages}{8418--8442}.
\newblock
\urldef\tempurl%
\url{https://aclanthology.org/2025.naacl-long.427/}
\showURL{%
\tempurl}


\bibitem[Sun et~al\mbox{.}(2025)]%
        {sun2025contextfolding}
\bibfield{author}{\bibinfo{person}{Weiwei Sun}, \bibinfo{person}{Miao Lu},
  \bibinfo{person}{Zhan Ling}, \bibinfo{person}{Kang Liu},
  \bibinfo{person}{Xuesong Yao}, \bibinfo{person}{Yiming Yang}, {and}
  \bibinfo{person}{Jiecao Chen}.} \bibinfo{year}{2025}\natexlab{}.
\newblock \bibinfo{title}{Scaling Long-Horizon LLM Agent via Context-Folding}.
\newblock
\showeprint[arxiv]{2510.11967}~[cs.CL]
\urldef\tempurl%
\url{https://arxiv.org/abs/2510.11967}
\showURL{%
\tempurl}


\bibitem[{SWE-Milestone Team}(2026)]%
        {swemilestone2026gozero}
\bibfield{author}{\bibinfo{person}{{SWE-Milestone Team}}.}
  \bibinfo{year}{2026}\natexlab{}.
\newblock \bibinfo{title}{{SWE-Milestone} go-zero M001: go-redis v9 Upgrade
  with API Modernization}.
\newblock \bibinfo{howpublished}{Official benchmark task specification}.
\newblock
\urldef\tempurl%
\url{https://huggingface.co/datasets/DeepCommit-ai/SWE-Milestone-data/blob/3500478e740d42583277c2658aa7729eb36e10f4/zeromicro_go-zero_v1.6.0_v1.9.3/srs/M001/SRS.md}
\showURL{%
\tempurl}


\bibitem[Wang et~al\mbox{.}(2025)]%
        {wang2025openhands}
\bibfield{author}{\bibinfo{person}{Xingyao Wang}, \bibinfo{person}{Boxuan Li},
  \bibinfo{person}{Yufan Song}, \bibinfo{person}{Frank~F. Xu},
  \bibinfo{person}{Xiangru Tang}, \bibinfo{person}{Mingchen Zhuge},
  \bibinfo{person}{Jiayi Pan}, \bibinfo{person}{Yueqi Song},
  \bibinfo{person}{Bowen Li}, \bibinfo{person}{Jaskirat Singh},
  {et~al\mbox{.}}} \bibinfo{year}{2025}\natexlab{}.
\newblock \showarticletitle{OpenHands: An Open Platform for AI Software
  Developers as Generalist Agents}. In \bibinfo{booktitle}{\emph{International
  Conference on Learning Representations}}.
\newblock
\urldef\tempurl%
\url{https://proceedings.iclr.cc/paper_files/paper/2025/hash/a4b6ad6b48850c0c331d1259fc66a69c-Abstract-Conference.html}
\showURL{%
\tempurl}


\bibitem[Wang et~al\mbox{.}(2026a)]%
        {wang2026swepruner}
\bibfield{author}{\bibinfo{person}{Yuhang Wang}, \bibinfo{person}{Yuling Shi},
  \bibinfo{person}{Mo Yang}, \bibinfo{person}{Rongrui Zhang},
  \bibinfo{person}{Shilin He}, \bibinfo{person}{Heng Lian},
  \bibinfo{person}{Yuting Chen}, \bibinfo{person}{Siyu Ye},
  \bibinfo{person}{Kai Cai}, {and} \bibinfo{person}{Xiaodong Gu}.}
  \bibinfo{year}{2026}\natexlab{a}.
\newblock \bibinfo{title}{{SWE-Pruner}: Self-Adaptive Context Pruning for
  Coding Agents}.
\newblock
\showeprint[arxiv]{2601.16746}~[cs.SE]
\urldef\tempurl%
\url{https://arxiv.org/abs/2601.16746}
\showURL{%
\tempurl}


\bibitem[Wang et~al\mbox{.}(2026b)]%
        {wang2026ledger}
\bibfield{author}{\bibinfo{person}{Zehao Wang}, \bibinfo{person}{Yisen Xu},
  \bibinfo{person}{Chenglin Li}, \bibinfo{person}{Chao Peng},
  \bibinfo{person}{Bram Adams}, \bibinfo{person}{Ahmed~E. Hassan}, {and}
  \bibinfo{person}{Tse-Hsun Chen}.} \bibinfo{year}{2026}\natexlab{b}.
\newblock \bibinfo{title}{Turning Interaction History into Execution State: A
  Runtime Layer for Long-Horizon Coding Agents}.
\newblock
\showeprint[arxiv]{2608.00808}~[cs.SE]
\urldef\tempurl%
\url{https://arxiv.org/abs/2608.00808}
\showURL{%
\tempurl}


\bibitem[{xAI}(2026)]%
        {xai2026grok46}
\bibfield{author}{\bibinfo{person}{{xAI}}.} \bibinfo{year}{2026}\natexlab{}.
\newblock \bibinfo{title}{{Grok} 4.6}.
\newblock \bibinfo{howpublished}{xAI API pricing documentation}.
\newblock
\urldef\tempurl%
\url{https://docs.x.ai/developers/pricing}
\showURL{%
\tempurl}


\bibitem[Xiao et~al\mbox{.}(2026)]%
        {xiao2026agentdiet}
\bibfield{author}{\bibinfo{person}{Yuan-An Xiao}, \bibinfo{person}{Pengfei
  Gao}, \bibinfo{person}{Chao Peng}, {and} \bibinfo{person}{Yingfei Xiong}.}
  \bibinfo{year}{2026}\natexlab{}.
\newblock \showarticletitle{Reducing Cost of LLM Agents with Trajectory
  Reduction}.
\newblock \bibinfo{journal}{\emph{Proceedings of the ACM on Software
  Engineering}} \bibinfo{volume}{3}, \bibinfo{number}{FSE}, Article
  \bibinfo{articleno}{FSE056} (\bibinfo{date}{July} \bibinfo{year}{2026}),
  \bibinfo{numpages}{22}~pages.
\newblock
\href{https://doi.org/10.1145/3797084}{doi:\nolinkurl{10.1145/3797084}}


\bibitem[Yang et~al\mbox{.}(2026b)]%
        {yang2026recursive}
\bibfield{author}{\bibinfo{person}{Chenxiao Yang}, \bibinfo{person}{Nathan
  Srebro}, {and} \bibinfo{person}{Zhiyuan Li}.}
  \bibinfo{year}{2026}\natexlab{b}.
\newblock \showarticletitle{Recursive Models for Long-Horizon Reasoning}. In
  \bibinfo{booktitle}{\emph{Proceedings of the 43rd International Conference on
  Machine Learning}} \emph{(\bibinfo{series}{Proceedings of Machine Learning
  Research}, Vol.~\bibinfo{volume}{306})}. \bibinfo{pages}{145939--145967}.
\newblock
\urldef\tempurl%
\url{https://proceedings.mlr.press/v306/yang26bp.html}
\showURL{%
\tempurl}


\bibitem[Yang et~al\mbox{.}(2024)]%
        {yang2024sweagent}
\bibfield{author}{\bibinfo{person}{John Yang}, \bibinfo{person}{Carlos~E.
  Jimenez}, \bibinfo{person}{Alexander Wettig}, \bibinfo{person}{Kilian
  Lieret}, \bibinfo{person}{Shunyu Yao}, \bibinfo{person}{Karthik Narasimhan},
  {and} \bibinfo{person}{Ofir Press}.} \bibinfo{year}{2024}\natexlab{}.
\newblock \showarticletitle{{SWE-agent}: Agent-Computer Interfaces Enable
  Automated Software Engineering}. In \bibinfo{booktitle}{\emph{Advances in
  Neural Information Processing Systems}}.
\newblock
\urldef\tempurl%
\url{https://proceedings.neurips.cc/paper_files/paper/2024/hash/5a7c947568c1b1328ccc5230172e1e7c-Abstract-Conference.html}
\showURL{%
\tempurl}


\bibitem[Yang et~al\mbox{.}(2026a)]%
        {yang2026programbench}
\bibfield{author}{\bibinfo{person}{John Yang}, \bibinfo{person}{Kilian Lieret},
  \bibinfo{person}{Jeffrey Ma}, \bibinfo{person}{Parth Thakkar},
  \bibinfo{person}{Dmitrii Pedchenko}, \bibinfo{person}{Sten Sootla},
  \bibinfo{person}{Emily McMilin}, \bibinfo{person}{Pengcheng Yin},
  \bibinfo{person}{Rui Hou}, \bibinfo{person}{Gabriel Synnaeve},
  \bibinfo{person}{Diyi Yang}, {and} \bibinfo{person}{Ofir Press}.}
  \bibinfo{year}{2026}\natexlab{a}.
\newblock \bibinfo{title}{{ProgramBench}: Can Language Models Rebuild Programs
  From Scratch?}
\newblock
\showeprint[arxiv]{2605.03546}~[cs.SE]
\urldef\tempurl%
\url{https://arxiv.org/abs/2605.03546}
\showURL{%
\tempurl}


\bibitem[Yao et~al\mbox{.}(2023)]%
        {yao2023react}
\bibfield{author}{\bibinfo{person}{Shunyu Yao}, \bibinfo{person}{Jeffrey Zhao},
  \bibinfo{person}{Dian Yu}, \bibinfo{person}{Nan Du}, \bibinfo{person}{Izhak
  Shafran}, \bibinfo{person}{Karthik Narasimhan}, {and} \bibinfo{person}{Yuan
  Cao}.} \bibinfo{year}{2023}\natexlab{}.
\newblock \showarticletitle{{ReAct}: Synergizing Reasoning and Acting in
  Language Models}. In \bibinfo{booktitle}{\emph{International Conference on
  Learning Representations}}.
\newblock
\urldef\tempurl%
\url{https://openreview.net/forum?id=WE_vluYUL-X}
\showURL{%
\tempurl}


\bibitem[Ye et~al\mbox{.}(2025)]%
        {ye2025kvcomm}
\bibfield{author}{\bibinfo{person}{Hancheng Ye}, \bibinfo{person}{Zhengqi Gao},
  \bibinfo{person}{Mingyuan Ma}, \bibinfo{person}{Qinsi Wang},
  \bibinfo{person}{Yuzhe Fu}, \bibinfo{person}{Ming-Yu Chung},
  \bibinfo{person}{Yueqian Lin}, \bibinfo{person}{Zhijian Liu},
  \bibinfo{person}{Jianyi Zhang}, \bibinfo{person}{Danyang Zhuo}, {and}
  \bibinfo{person}{Yiran Chen}.} \bibinfo{year}{2025}\natexlab{}.
\newblock \showarticletitle{KVCOMM: Online Cross-context KV-cache Communication
  for Efficient LLM-based Multi-agent Systems}. In
  \bibinfo{booktitle}{\emph{Advances in Neural Information Processing
  Systems}}, Vol.~\bibinfo{volume}{38}. \bibinfo{pages}{20359--20405}.
\newblock
\href{https://doi.org/10.52202/085713-0605}{doi:\nolinkurl{10.52202/085713-0605}}


\bibitem[Ye et~al\mbox{.}(2026)]%
        {ye2025agentfold}
\bibfield{author}{\bibinfo{person}{Rui Ye}, \bibinfo{person}{Zhongwang Zhang},
  \bibinfo{person}{Kuan Li}, \bibinfo{person}{Huifeng Yin},
  \bibinfo{person}{Zhengwei Tao}, \bibinfo{person}{Yida Zhao},
  \bibinfo{person}{Liangcai Su}, \bibinfo{person}{Liwen Zhang},
  \bibinfo{person}{Wenbiao Yin}, \bibinfo{person}{Zile Qiao},
  \bibinfo{person}{Xinyu Wang}, \bibinfo{person}{Pengjun Xie},
  \bibinfo{person}{Fei Huang}, \bibinfo{person}{Jingren Zhou},
  \bibinfo{person}{Siheng Chen}, {and} \bibinfo{person}{Yong Jiang}.}
  \bibinfo{year}{2026}\natexlab{}.
\newblock \showarticletitle{{AgentFold}: Long-Horizon Web Agents with Proactive
  Context Folding}. In \bibinfo{booktitle}{\emph{International Conference on
  Learning Representations}}.
\newblock
\urldef\tempurl%
\url{https://proceedings.iclr.cc/paper_files/paper/2026/hash/fa948624dfde013671e72c1a7ca4aebc-Abstract-Conference.html}
\showURL{%
\tempurl}


\bibitem[Young(2025)]%
        {young2025harnesses}
\bibfield{author}{\bibinfo{person}{Justin Young}.}
  \bibinfo{year}{2025}\natexlab{}.
\newblock \bibinfo{title}{Effective Harnesses for Long-Running Agents}.
\newblock \bibinfo{howpublished}{Anthropic Engineering}.
\newblock
\urldef\tempurl%
\url{https://www.anthropic.com/engineering/effective-harnesses-for-long-running-agents}
\showURL{%
\tempurl}


\bibitem[Zeng et~al\mbox{.}(2026)]%
        {zeng2026attncompress}
\bibfield{author}{\bibinfo{person}{Zhengran Zeng}, \bibinfo{person}{Yixin Li},
  \bibinfo{person}{Rui Xie}, \bibinfo{person}{Wei Ye}, {and}
  \bibinfo{person}{Shikun Zhang}.} \bibinfo{year}{2026}\natexlab{}.
\newblock \showarticletitle{{AttnCompress}: Dynamic Attention-Guided Trajectory
  Compression for Software Engineering Agents}.
\newblock \bibinfo{journal}{\emph{Proceedings of the ACM on Software
  Engineering}} \bibinfo{volume}{3}, \bibinfo{number}{ISSTA}
  (\bibinfo{date}{Oct.} \bibinfo{year}{2026}), \bibinfo{pages}{1331--1353}.
\newblock
\href{https://doi.org/10.1145/3832149}{doi:\nolinkurl{10.1145/3832149}}


\bibitem[Zhang et~al\mbox{.}(2025)]%
        {zhang2025agentprune}
\bibfield{author}{\bibinfo{person}{Guibin Zhang}, \bibinfo{person}{Yanwei Yue},
  \bibinfo{person}{Zhixun Li}, \bibinfo{person}{Sukwon Yun},
  \bibinfo{person}{Guancheng Wan}, \bibinfo{person}{Kun Wang},
  \bibinfo{person}{Dawei Cheng}, \bibinfo{person}{Jeffrey Yu}, {and}
  \bibinfo{person}{Tianlong Chen}.} \bibinfo{year}{2025}\natexlab{}.
\newblock \showarticletitle{Cut the Crap: An Economical Communication Pipeline
  for LLM-based Multi-Agent Systems}. In
  \bibinfo{booktitle}{\emph{International Conference on Learning
  Representations}}.
\newblock
\urldef\tempurl%
\url{https://proceedings.iclr.cc/paper_files/paper/2025/file/bbc461518c59a2a8d64e70e2c38c4a0e-Paper-Conference.pdf}
\showURL{%
\tempurl}


\bibitem[Zhang et~al\mbox{.}(2026)]%
        {zhang2026autocompact}
\bibfield{author}{\bibinfo{person}{Xuan Zhang}, \bibinfo{person}{Longtao
  Zheng}, \bibinfo{person}{Cunxiao Du}, \bibinfo{person}{Bo An}, {and}
  \bibinfo{person}{Xin Dong}.} \bibinfo{year}{2026}\natexlab{}.
\newblock \bibinfo{title}{{AutoCompact}: Learning When to Compact Context in
  Long-Horizon Coding Agents}.
\newblock
\showeprint[arxiv]{2610.02163}~[cs.CL]
\urldef\tempurl%
\url{https://arxiv.org/abs/2610.02163}
\showURL{%
\tempurl}


\bibitem[Zheng et~al\mbox{.}(2024)]%
        {zheng2024sglang}
\bibfield{author}{\bibinfo{person}{Lianmin Zheng}, \bibinfo{person}{Liangsheng
  Yin}, \bibinfo{person}{Zhiqiang Xie}, \bibinfo{person}{Chuyue Sun},
  \bibinfo{person}{Jeff Huang}, \bibinfo{person}{Cody~Hao Yu},
  \bibinfo{person}{Shiyi Cao}, \bibinfo{person}{Christos Kozyrakis},
  \bibinfo{person}{Ion Stoica}, \bibinfo{person}{Joseph~E. Gonzalez},
  \bibinfo{person}{Clark Barrett}, {and} \bibinfo{person}{Ying Sheng}.}
  \bibinfo{year}{2024}\natexlab{}.
\newblock \showarticletitle{{SGLang}: Efficient Execution of Structured
  Language Model Programs}. In \bibinfo{booktitle}{\emph{Advances in Neural
  Information Processing Systems}}, Vol.~\bibinfo{volume}{37}.
\newblock
\urldef\tempurl%
\url{https://proceedings.neurips.cc/paper_files/paper/2024/hash/724be4472168f31ba1c9ac630f15dec8-Abstract-Conference.html}
\showURL{%
\tempurl}


\bibitem[Zheng et~al\mbox{.}(2026)]%
        {zheng2026corvus}
\bibfield{author}{\bibinfo{person}{Mingwei Zheng}, \bibinfo{person}{David
  O'Brien}, \bibinfo{person}{Siwei Cui}, \bibinfo{person}{Pardis Pashakhanloo},
  \bibinfo{person}{Rajdeep Mukherjee}, \bibinfo{person}{Myeongsoo Kim}, {and}
  \bibinfo{person}{Sachit Kuhar}.} \bibinfo{year}{2026}\natexlab{}.
\newblock \bibinfo{title}{{CORVUS}: Context Optimization and Reduction Via
  Underlying Synchronization for {LLM} Coding Agents}.
\newblock
\showeprint[arxiv]{2607.22711}~[cs.LG]
\urldef\tempurl%
\url{https://arxiv.org/abs/2607.22711}
\showURL{%
\tempurl}


\bibitem[Zhu et~al\mbox{.}(2025)]%
        {zhu2025multiagentbench}
\bibfield{author}{\bibinfo{person}{Kunlun Zhu}, \bibinfo{person}{Hongyi Du},
  \bibinfo{person}{Zhaochen Hong}, \bibinfo{person}{Xiaocheng Yang},
  \bibinfo{person}{Shuyi Guo}, \bibinfo{person}{Zhe Wang},
  \bibinfo{person}{Zhenhailong Wang}, \bibinfo{person}{Cheng Qian},
  \bibinfo{person}{Xiangru Tang}, \bibinfo{person}{Heng Ji}, {and}
  \bibinfo{person}{Jiaxuan You}.} \bibinfo{year}{2025}\natexlab{}.
\newblock \showarticletitle{MultiAgentBench: Evaluating the Collaboration and
  Competition of LLM agents}. In \bibinfo{booktitle}{\emph{Proceedings of the
  63rd Annual Meeting of the Association for Computational Linguistics (Volume
  1: Long Papers)}}. \bibinfo{publisher}{Association for Computational
  Linguistics}, \bibinfo{pages}{8580--8622}.
\newblock
\href{https://doi.org/10.18653/v1/2025.acl-long.421}{doi:\nolinkurl{10.18653/v1/2025.acl-long.421}}


\end{thebibliography}

\end{document}